\documentclass[11pt,a4paper]{article}

\usepackage{fancyhdr}

\usepackage{hyperref}
\usepackage{graphics}
\usepackage{multirow}
\usepackage{graphicx,epsfig,color}
\usepackage{epstopdf}
\usepackage{url}
\usepackage{units}
\usepackage{amssymb,amsmath,array,wasysym}
\usepackage{lscape}
\usepackage{colortbl}
\usepackage{booktabs}

\DeclareFontFamily{U}{euc}{}
\DeclareFontShape{U}{euc}{m}{n}{<-6>eurm5<6-8>eurm7<8->eurm10}{}%
\DeclareSymbolFont{AMSc}{U}{euc}{m}{n} 
\DeclareMathSymbol{\umu}{\mathord}{AMSc}{"16}

\begin{document}

\title{Summary of the Linear Collider Testbeam Workshop 2009 \\ LCTW09}

\author{
V.~Boudry$^1$, G.~Fisk$^2$, R.E.~Frey$^3$, F.~Gaede$^4$, C.~Hast$^5$, J.~Hauptman$^{6}$, \\
K.~Kawagoe$^7$, L.~Linssen$^8$, R.~Lipton$^2$, W.~Lohmann$^9$, T.~Matsuda$^{3, 10}$, \\
T.~Nelson$^6$,  R.~P\"oschl$^{11}$, E.~Ramberg$^2$, F.~Sefkow$^4$, M.~Vos$^{12}$, M.~Wing$^{13}$, 
J.~Yu$^{14}$\\
{\footnotesize {\it 1- Laboratoire Leprince-Ringuet  (LLR), \'Ecole Polytechnique - CNRS/IN2P3}}\\
{\footnotesize {\it Route de Saclay, 91128 Palaiseau Cedex, France}}\\
{\footnotesize {\it 2- FNAL, P.O. Box 500, Batavia, IL, 60510-0500, USA}}\\
{\footnotesize {\it 3- Physics Department, 1274 University of Oregon, Eugene, OR 97403, USA}}\\
{\footnotesize {\it 4- DESY, Notkestrasse 85, D-22603 Hamburg, Germany}}\\
{\footnotesize {\it 5- SLAC, 2575 Sand Hill Road, Menlo Park, CA 94025, USA}}\\
{\footnotesize {\it 6- Department of Physics and Astronomy, 12 Physics Hall, Ames, IA 50011, USA}}\\
{\footnotesize {\it 7- Department of Physics, Kobe University, Kobe, 657-8501, Japan}}\\
{\footnotesize {\it 8 CERN, 1211 Gen\`{e}ve 23, Switzerland}}\\
{\footnotesize {\it 9- DESY, Platanenall\'ee 6, D-15738 Zeuthen, Germany}}\\
{\footnotesize {\it 10- KEK, 1-1 Oho, Tsukuba Ibaraki 305-0801, Japan}}\\
{\footnotesize {\it 11- Laboratoire de l'Acc\'el\'erateur Lin\'eaire (LAL) -CNRS/IN2P3; B.P. 34, 91898 Orsay Cedex, France}}\\
{\footnotesize {\it 12- IFIC, Centro Mixto CSIC-UVEG, Edificio Investigacion Paterna, Apartado 22085, }}\\
{\footnotesize {\it 46071 Valencia, Spain}}\\
{\footnotesize {\it 13- Department of Physics and Astronomy, University College London, Gower Street,}}\\
{\footnotesize {\it  London WC1E 6BT, UK}}\\
{\footnotesize {\it 14- Department of Physics, SH108, University of Texas, Arlington, TX 76019, USA}}\\
}

\date{}
\maketitle
\thispagestyle{fancy}


\begin{abstract}
This note summarises the workshop LCTW09 held between the 3.11.2009 and 5.11.2009 at LAL Orsay. The workshop was dedicated to discuss the beam tests in the years 2010 up to 2013 for detectors to be operated at a future linear electron positron collider. The document underlines the
rich R\&D program on these detectors in the coming years. 
Large synergies were identified in the DAQ and
software systems. Considerable consolidation of resources are expected from the establishment of semi-permanent beam lines for
linear collider detector R\&D at major centres like CERN and FNAL. Reproducing a beam structure as foreseen for the International Linear Collider (ILC) would clearly enhance the 
value of the obtained beam test results. Although not ultimately needed for every research program, all groups would exploit such a feature if it is available. 


\end{abstract}



\section{Executive summary}

The next major worldwide project in high energy particle
physics will be a linear electron positron collider at the TeV scale. This machine will complement and extend
the scientific results of the LHC currently operated at CERN. The most advanced proposal for such a machine is the {\em I}nternational
{\em L}inear {\em C}ollider (ILC) which will be operated at centre-of-mass energies between 90\,GeV and 1\,TeV. The experimentation at this machine could start around the year 2020.  Introductions to  the concept of the ILC and to the alternative for higher energies CLIC, short for {\em C}ompact {\em LI}near {\em C}ollider concept, can be found elsewhere~\cite{rdr07, clic04}. 

Beams tests are necessary for detector concepts to validate their design and gain valuable operating experience. At the same time
they are an optimal opportunity to train young physicist on real data. In November 2009, 40 experts
(two from Asia, five from North America and the rest from Europe) met at the Laboratoire
de l'Acc\'el\'erateur Lin\'eaire (LAL) at Orsay to review the needs for future beam tests for the R\&D on detectors~\cite{lctw09}.
The goal of this workshop was to collect the needs and to coordinate the activities of the various collaborations
active in the field: CALICE, FCAL, SiD as well as groups working on the dual readout technology on calorimetry, LCTPC on gaseous tracking as well as SiLC for the
various silicon tracking devices. Representatives of the current major test beam facilities, CERN, DESY and Fermilab,
presented their sites and actively took part in the discussions. Many other facilities available in the world were discussed:
J-Parc, IHEP Bejing, Tohoku, KEK in Asia, IHEP/Protvino, Dubna in Russia, and it was noticed that SLAC would restore test
beams and create a new facility in its End Station A by 2010. The successful beam test efforts prior to the Letters of Intent
(LOI) for detectors~\cite{ild-loi, sid-loi, fourth_LOI} were reviewed followed by discussions on what is needed to improve these test beams for the next
phase. This document covers the years 2010--2013 which to a large extent coincides with the preparation of the Detector Baseline
Document (DBD) in which mature detector technologies are to be presented. The beam test efforts have to support this goal. Large scale
systems of all detector components are expected to be tested in this phase. The successful conduction of the beam tests is naturally 
vital for a well founded document. Apart from the fact that the detector developers have to be ready in time, the
community has to make sure that enough beam time is available, in particular in the period 1/2011 - 2/2012 in which most of 
the activities described in this note can be anticipated. The needs of the linear collider detector community in terms of particles comprises 
high and low energy electron as well as high and low energy hadron beams. In addition to the beam lines, the linear collider 
detector R\&D requires specific equipment such as large-bore high-field magnets (up to 6\,T). The Table~\ref{tab:testbeams} gives a general overview on the activities planned by the various detector components.
Another important issue of the detector R\&D is to find the optimal balance between high beam rates to conduct physics motivated studies and the fact that e.g. the readout electronics is primarily designed for low rates as in initial operation expected at a linear collider. 
While the priority is clearly to be given to the availability of test beams, the establishment of a dedicated ILC beam structure would render the results more applicable to prospects on the operation at the ILC. This would be particularly true for the general time structure, i.e. ``macro-structure'' of the beam. This means that a relatively short pulse of about 1\,ms is
followed by a longer interval of up to 199\,ms without beam. All the R\&D groups would make use of such a possibility to test their hardware under 
the most realistic conditions. There would be no strong requirement to reproduce the micro-structure of the ILC beams. 
The community encourages the site operators to continue efforts to establish an ILC like beam structure.
Given the limited time line and manpower situation the LC Detector community will establish a light coordination of 
the beam test activities to foster synergies and avoid overlaps in terms of beam times and facilities.

The activities of the past and the challenging program of the future have been acknowledged in Europe by the recent approval of the {\em AIDA} project. 
The AIDA project~\cite{aida} was presented as a proposal for Integrating Activity under the 7th framework of the European Union. This funding instrument 
is largely based on the ``I3'' (integrated infrastructure initiatives) under which EUDET~\cite{eudet} was funded. The objectives are (a) to provide a wider 
and more efficient access to, and use of the existing research infrastructures in Europe and (b) better integration of the way research infrastructures 
operate, and fostering joint development in terms of capacity and performance. AIDA involves most of the linear collider detector R\&D community that 
participated in EUDET. It moreover includes a strong participation from the institutes involved in the upgrades of the LHC detectors, in that of the 
B-factories and accelerator-based neutrino experiments. At the time of writing, AIDA was in the negotiation phase after the EU had proposed to fund 
the proposal with 8 million euro (compared to the 10 million requested from the EU in the original proposal). In the proposal common infrastructure 
for the characterisation of new detector prototypes is foreseen. It should be emphasised that the programs were and are open for participation of non-european
partners in the actual projects.

\begin{landscape}
\begin{table}[htdp]
\begin{center}
\begin{tabular}{@{} |c|cc|cc|cc|cc|cc| @{}}
 \hline
    Project  & 2010/2 & Site& 2011/1 & Site & 2011/2 & Site & 2012/1 & Site & 2012/2 & Site\\
    \hline
    Calo& LS &  {\bf CERN}  & LS & {\bf CERN} & LS & {\bf CERN} & LS &  {\bf CERN} & LS &  {\bf CERN}\\
        &    & {\bf FNAL} &    & {\bf FNAL} &    &   {\bf FNAL}  &    &  {\bf FNAL} &    & {\bf FNAL}\\
        &    &      SLAC  &    & {\bf SLAC} &    &   {\bf SLAC}  &    &       SLAC  &    &     SLAC\\
   Needs& \multicolumn{2}{c|}{Magnet}   & \multicolumn{2}{c|}{Magnet} & \multicolumn{2}{c|}{Magnet}& \multicolumn{2}{c|}{Magnet}& \multicolumn{2}{c|}{Magnet}\\
        &    & \multicolumn{9}{c|}{Particle types: $e$, $\pi$, $p$, energies: 1-120\,GeV, low rates $\approx 100$\,Hz}\\   
    \hline
    Gas/TPC& LS&                & LS & CERN & LS  &     CERN & LS  &  CERN & ? &      CERN \\
           &   &  {\bf DESY}    &    & {\bf DESY}  &    &   {\bf DESY}  &    &   {\bf DESY}  &   &   {\bf DESY}\\
           &   &                 &    &   FNAL   &    &   FNAL  &    &   FNAL  &   &   FNAL\\
   Needs& \multicolumn{2}{c|}{Magnet}   & \multicolumn{2}{c|}{Magnet} & \multicolumn{2}{c|}{Magnet}& \multicolumn{2}{c|}{Magnet}& \multicolumn{2}{c|}{Magnet}\\
        &    & \multicolumn{9}{c|}{Particle types and rates: $e$ as available at DESY, hadron beam test not planned but possible.}\\   
    \hline
    SiTrack& SU &    Various (see Tab.\ref{tab:sitr})&  SU &       Various  & SU  &   Various & SU &    Various & SU &      Various\\
    Needs& \multicolumn{2}{c|}{Magnet/Telescope}   & \multicolumn{2}{c|}{M./T.} & \multicolumn{2}{c|}{M./T.}& \multicolumn{2}{c|}{M./T.}& \multicolumn{2}{c|}{M./T.}\\
        &    & \multicolumn{9}{c|}{Particle types: $e$, $\pi$, $p$, energies: 1-120\,GeV, high rates $\approx 1$\,MHz for short periods}\\   
    \hline
\end{tabular}
\end{center}
\caption{\sl The table indicate the envisaged beam test activities until the end of 2012. The acronym {\bf SU} means ``Test of Small Units can be expected'', The acronym {\bf LS} means ``Large Scale Testbeam planned''. Sites are given in {\em alphabetical} order. Bold face letters indicate where beam tests are going to happen. Normal face letters indicate optional tests depending on availability of detector prototypes and needs. The required strength of the magnetic field is between 3 and 6\u{T}.}
\label{tab:testbeams}
\end{table}%
\end{landscape}

\section{World wide linear collider beam test coordination and review}

Efforts on the International Linear Collider increased after the creation of the 
Global Design Effort (GDE) in 2004.   Along with the global effort on the accelerator front, many detector development 
groups intensified their activities and their need for beam tests rose.  These efforts were, however,  
fragmented and not coordinated. Given the anticipated intensity of beam test efforts in the coming few 
years, it was necessary for the community and the facilities to be able to provide necessary beam capabilities to 
detector R\&D groups. The facilities, however, needed to know what the requirements for the community are.  
As an effort to convey the upcoming needs, the calorimeter and muon R\&D groups have put together a road map document 
to FNAL in 2005, following a presentation to the Physics Advisory Panel (PAC). This document~\cite{road05} and the need for 
more concerted effort led to the implementation of a working group structure and prompted the need for a world-wide 
ILC test beam workshop to collect and compile the requirements of most, if not all, R\&D groups within the community.    
This was to provide a forum to share ideas and needs between many different groups within the LC community and to make sure 
that the limited facilities can be used effectively.

\subsection{LC Test Beam Workshop 2007 (IDTB07) at FNAL}

The LCTW09 as summarised in this document is the second workshop of this kind. 
The first workshop on linear collider beam tests, called IDTB07, was held at FNAL in Jan. 2007.

As a result of three days of presentations and discussions at IDTB07, the following requirements were identified:
\begin{itemize}
\item Large bore, high field magnet (up to 5T);
\item ILC beam time structure (1ms beam + 199ms blank); 
\item Mimicking hadron jets;
\item Common DAQ hardware and software;
\item Common online and offline software;
\item Common reconstruction and analysis software infrastructure;
\item Tagged neutral hadron beams.
\end{itemize}

The outcome of the IDTB07 workshop resulted in a roadmap document~\cite{idtb07} that was released to 
the linear collider leadership and facility managers in summer 2007.  Many of the improvements made in facilities in subsequent years were based on the requirements and the roadmap laid down in this document. Among the efforts to support the linear collider detector R\&D, the following are to be highlighted:
\begin{itemize}
\item The CALICE collaboration benefited from the availability of the H6 beam test area at CERN over several months in the years 2006 and 2007. This considerable beam time was allocated on short notice, despite the huge demands required by the final
stages of the LHC detector R\&D program and the launch of the neutrino program at CERN.
\item FNAL refurbished the MTest beam line particularly to host the CALICE beam test program in the years 2008 and 2009. This program is to be pursued in 2010 and beyond. The continuing availability of the test beam facility at both CERN and FNAL allowed
the establishment of an infrastructure by which CALICE was able to setup remote control facilities which are a first step towards
a similar detector control at a future linear collider.
\item The DESY facility gave a 'home' to the TPC activities which could then establish the infrastructure needed to pursue the R\&D at 
a single place. 
\item Beyond the activities above, the various sites, i.e. CERN, DESY, FNAL, SLAC and KEK, offered beam time to smaller yet very important activities by the vertex, silicon tracking and muon detector communities.
\end{itemize}
{\em The detector R\&D community would like to take the opportunity here to express their acknowledgement and gratitude for the effort made by the test beam sites.}

The beam test activities resulted in a number of scientific results which can be viewed on the web-pages of the different projects.

\section{Subdetector beam test plans}
\subsection{Calorimeter}
\label{calo}
As will be outlined in this section the calorimeters may put the highest
demands in terms of space and availability of beam test areas.  %
Many projects feature prototypes of about 1\u{m^3} and need high statistics for
the conduction of physics programs during the beam test campaigns.


\subsubsection{CALICE plans}

An overview of past, present and future CALICE calorimeter prototypes is
available in Table~\ref{tab:calicecals}.  %
For details on the CALICE program, the reader is referred
to~\cite{calice-prc2009}.
\definecolor{light-gray}{gray}{0.70}
\begin{table}[htdp]
\begin{center}
\begin{footnotesize}
\begin{tabular}{@{} |rlccc| @{}}
  \hline
  Project 	          & Type 			            & Absorber 	                   & Sensitive part	    & Completion date     \\
  \hline
  \multirow{2}*{AHCAL} 	  & Physics prototype  	                    & Stainl. steel                & Scintillator 	    & Completed           \\
  \arrayrulecolor{light-gray} \cline{2-5} \arrayrulecolor{black} 
  ~                       & Technological prototype                 & Stainl. steel                & Scintillator 	    & 2012	          \\
  \hline
  TCMT 			  & Physics prototype   	            & Stainl. steel                & Scintillator 	    & Completed           \\
  \hline
  \multirow{2}*{DHCAL} 	  & \multirow{2}*{Physics prototype}        & \multirow{2}*{Stainl. steel} & RPC 		    & \multirow{2}*{2010} \\
  ~                       &                                         &                              & Partially GEM 	    &                     \\ 
  \arrayrulecolor{light-gray} \hline \arrayrulecolor{black} 
  \multirow{2}*{SDHCAL}   & \multirow{2}*{Physics \& Technological} & \multirow{2}*{Stainl. steel} & RPC 		    & \multirow{2}*{2011} \\
  ~                       &                                         &                              & Partially $\umu$Megas  &                     \\
  \hline
  \multirow{3}*{W HCAL}   & \multirow{3}*{Physics prototype}        & \multirow{3}*{Tungsten} 	   & Scintillator 	    & \multirow{3}*{2011} \\
  ~                       &                                         &                              & Partially $\umu$Megas  &                     \\
  ~                       &                                         &                              & Partially GEM	    &                     \\
  \hline
  \multirow{3}*{SiW Ecal} & Physics prototype                       & Tungsten	                   & Si 		    & Completed           \\
  \arrayrulecolor{light-gray} \cline{2-5} \arrayrulecolor{black} 
  ~                       & \multirow{2}*{Technological prototype}  & \multirow{2}*{Tungsten}	   & Si 		    & \multirow{2}*{2012} \\
  ~                       &                                         &                              & Partially scintillator &                     \\
  \hline
  ScW Ecal                & Physics prototype	                    & Tungsten 	                   & Scintillator 	    & Completed           \\ 
  \hline
  DECAL                   & Physics prototype 	                    & Tungsten 	                   & Si 		    & ?               \\
  \hline
\end{tabular}
\end{footnotesize}
\end{center}
\caption{\sl Overview of calorimeter prototypes having been or to be operated by the CALICE collaboration. }
\label{tab:calicecals}
\end{table}

Each project has developed or is developing prototype(s) classified as
\emph{physics}, used to demonstrate the physics performances of the technique,
or \emph{technological}, used to study the solutions to the technological
constraints arising from the integration in a large ILC
detector\footnote{mainly heating, mechanical integration, compactness, embedded
  FE electronics, power-pulsing}, or both.  %
The Digital ECAL (DECAL) technique has only been studied in
small devices and no full-scale physics prototype has been built.
As this project depended on UK funding which has now been cancelled,
it is unlikely such a prototype will be constructed in the
foreseeable future.

Two generation of DAQ system have been developed: the first version, more
specifically dedicated to physics prototypes, has been running for a few
years. %
The second version, suited for technological prototypes and handling the
readout of a large quantity of channels digitised in the detectors, is at the
end of its development phase. More details are given in Section~\ref{sec:daq}.

\paragraph{Physics prototypes} 

The years 2010--11 will see the finalisation of the main physics prototype
phase. %
A physics prototype of a digital hadron calorimeter (DHCAL) based on thin RPCs
and $1\times1\u{cm^2}$ cells, will be completed in the first half of 2010.  
As for previous beam tests including the analogue hadron calorimeter (AHCAL),
besides standalone data taking, there will be data taking in combination with
the physics prototype of the electromagnetic silicon tungsten calorimeter (SiW
ECAL) and the Tail Catcher and Muon tracker (TCMT).  

Including commissioning and calibration phases altogether, 14 weeks of test
beam time will be requested from FNAL.  %
Within these 14 weeks, CALICE should be the primary beam user for about 8
weeks.  %
The other 6 weeks are devoted to the setup of the experiment in parasitic
running mode.  %
The physics program to be conducted is largely similar to the corresponding
data taking in the years 2006--09 with the AHCAL. 
In the combined running, the emphasis will be put on energy ranges in which it
is expected to see signals in the electromagnetic part and the hadronic part
(plus tail catcher).  
In the standalone running low energy hadrons and electrons are also to be
collected.  Priorities will have to be defined later on but the data which were
already taken give good guidelines.  
It is also envisaged to replace a few layers of the DHCAL with GEMs as
sensitive detectors.  This may happen in 2011.  

A new initiative, dubbed W-HCAL, has been started within CALICE in order to
study the properties of tungsten as absorber material, primarily for a compact
HCAL at a multi-TeV collider.  
A versatile structure, featuring forty 16\u{mm}-thick tungsten-alloy absorbers,
is foreseen.  %
Tests with existing scintillator layers are planned for end of 2010 and 2011 at
CERN. Tests with gaseous insert and second generation scintillator as they become available.

\paragraph{Technological prototypes}

The CALICE collaboration is entering a new phase of R\&D in which readout
technologies and mechanical designs meet many requirements of the
operation in a detector for a Linear Collider.  
Several groups of the collaboration are already quite advanced and new full
scale prototypes are expected towards the end of 2010.  The finalisation of
these prototypes will be preceded by a number of smaller beam test
efforts which will allow for maturing the newly developed technologies.
Examples for these test beam efforts are:
\begin{itemize} 
\item Beam tests with 1\u{m^2} units of the technical prototype of the SDHCAL
  (both RPC and $\umu$Megas variants).
\item The AHCAL conducted an initial small scale beam test at the beginning of
  2010 to prepare for electronics commissioning, to be followed by a so-called
  horizontal test towards the end of 2010 and a vertical test in 2011.  This
  means the available equipment will be arranged to allow for the measurement
  of electromagnetic showers.
\item The Si-W ECAL is planning to make tests with single ASU towards the
  beginning of 2011 in a beam test with electrons.  %
\end{itemize}

It has to be stressed that the primary goal of these prototypes is to study
technological solutions for the calorimetry at the ILC.  
The strategy for the coming years should take this into account. 
Here the main keywords are power pulsing, with a duty-cycle of typically
1\u{\%}, and limited depth of the buffers in the front end electronics.  
Hence the provided particle rates should not exceed 1\u{kHz} during a
spill.  
This is even more limited for RPCs, due to their comparatively large recovery
time, requiring rates $\lesssim 0.1\u{kHz}$.  %

In addition to the pure technological issues a physics program is to be
pursued.  %
Derived from those of the physics prototypes, taking the technical constraints
into account, it requires the operators of beam test sites to actively respond
to the needs of the CALICE (LC) beam test data taking at an very early stage.  %
As it is foreseeable that potential high statistics physics runs will take a
considerable amount of time, this will require the deployment of remote control
at the experimental sites.  
As some prototypes may use flammable gas, the topic of safety will have to be
addressed at a very early stage.

A first large scale beam test with a fully equipped technical prototype of an
SDHCAL can be expected towards spring 2011.  %
It is still to be clarified in what proportion this cubic meter prototype will
be equipped with the two technologies under study, namely using Glass RPCs or
$\umu$Megas as sensitive devices.  %
This is currently being reviewed on the basis of experience gained with the two
technologies by laboratory studies and during test beam campaigns of the year
2009.

Ideally, the SDHCAL will be joined by an SiW ECAL technological prototype by
the end of 2011.  %
The running of an AHCAL technical prototype alone and together with the SiW
ECAL technical prototype is to follow.  %
During the year 2010 mechanical interfaces between the different detector types
will have to be defined.  %
More generally the year 2010 is to be used to integrate the detector components
with the newly developed DAQ systems in order to provide an efficient data
taking.

\paragraph{}

The program requires a high availability of beam test areas.  
The CALICE management and the CALICE TB together with the corresponding ILC
R\&D panels will work out until summer 2010 whether ILC detector R\&D can
occupy consecutively beam test areas for a time of two or more years starting
with the beginning of 2011.  Such a high availability of beam test areas would
also allow for an easier conduction of smaller beam test programs as for example
with the DECAL. %
In addition the infrastructure would facilitate the testing of a prototype for
the electromagnetic calorimeter based on scintillating tiles (ScW ECAL) of
which one layer can be expected towards the end of 2012.  Finally,
technological prototype layers with timing capabilities should also be used in
a beam test with a tungsten absorber structure.

\subsubsection{SiD ECAL}
\label{sec:SiD-ECAL}

The silicon tungsten ECAL developed specifically for SiD features 30
longitudinal sampling layers composed of hexagonal high resistivity silicon
wafers divided in small hexagonal cells (13\u{mm^2}).  %
The readout of 1024 channels is performed by a single KPiX chip bump-bonded
directly on the wafer.  %
The chip is connected to the DAQ by flat polyimide cables.  %
The R\&D on components is almost completed and a compact stack prototype (30
layers of one wafers, interleaved with $15\times15\u{cm^2}$ tungsten alloy
absorbers) is being built and should be ready for test beam in beginning of
2011.

The ideal test beam for initial test is a 5--10\u{GeV} (or more) electron beam,
well localised and controllable, with a LC-like time structure (for KPiX
electronics).  %
A small number of electrons (mean of $\sim1-2$) per bunch is a must. Such a
beam is possibly available at SLAC, with a low rate ($< 60\u{Hz}$).

The data taking is planned for 2011, preferably at SLAC if a beam exists by
then (the current expectation is to have SLAC test beam available around winter
2011).  The possibility to realise combined tests with a HCAL prototype with a
hadron beam in 2012--13, needs to be evaluated.

\subsubsection{SiD Muons}
\label{sec:Muons}

The muon system of the SiD concept~\cite{sid-loi} will be placed after a thin
(5\u{\lambda_I}) calorimeter, the solenoid coil and cryostat (1.3\u{\lambda_I})
and is therefore crucial to measure leakage of highly energetic and
late-developing showers.  %
It features a total detector area of about 6000\u{m^2} on 14 layers for a total
number of channels of 50K (if single ended) to 100K (if dual readout).

The main criteria of choice are the cost, the ease of shape adaptation and
performance and reliability.  
The need to operate inside the return yoke adds the following: insensitivity to
magnetic field, space economy for the readout system (cables, FE, etc),
reliability and slow ageing.

Two technologies are considered:
\begin{description}
\item[RPC based (baseline)] Uses a variant of the KPiX readout chip, double gap
  Bakelite RPCs operating in the avalanche mode are or will be tested.  This
  effort benefits from synergy with the DHCAL (readout ASIC) and a long
  experience in various experiments (BaBar, Opera, BES-III, \dots) but some
  ageing and reliability issues have still to be clarified, using cosmic ray
  test stands with radioactive sources for tests of new Beijing Bakelite
  manufactured RPCs for used in the BES III and Daya Bay muon systems. %
  Since RPCs using this material have achieved acceptable dark noise rates
  without linseed oil coating, these RPCs look promising~\cite{mu-rpc}, but
  ageing effects have not been thoroughly studied.  %

\item[Scintillator based (alternative)]: wave-length-shifting fibre readout of
  cheap extruded scintillator coupled with new (and potentially) low-cost
  Si-based avalanche photo-diodes (Pixelated Photon Detectors or PPDs, also
  dubbed SiPMs) make the scintillator alternative progressively more
  competitive if long strips of up to 6\u{m} are feasible. %
  Prototypes featuring 256 scintillator strips and Multi Anode PMTs have been
  tested in the Fermilab Meson Test Beam Facility (MTBF), see
  Section~\ref{fnal}.  %
  A small sample of prototype were tested in 2008 at the Fermilab MTBF with a
  new type of SiPM, developed by FBK-IRST (Trento, Italy) to match the circular
  cross section of WLS fibre and have also shown good results.  %
  The main objectives of the MTBF tests have been to measure an expected
  increase in number of photo-electrons due to the increased quantum efficiency
  of the PPDs and to use the spontaneous release of photo-electrons that is
  normally called ``noise'' to measure the gain vs.\ bias voltage and thereby
  establish a calibration without special additional equipment. \\
  For an overview on the results of the tests described before see
  e.g.~\cite{mu-ppd}.
\end{description}

The short term plans are for RPCs at SLAC to be read out using multiple KPiX
chips in cosmic-ray test stands with BaBar and Chinese RPCs (Henry Band).  %
In a separate parallel cosmic ray setup at Princeton, Chang-guo Lu and
colleagues will test new Chinese RPCs manufactured commercially.  %
A major objective of the cosmic ray tests is the measurement of efficiency in a
high background artificially created with radioactive sources.

The prototype scintillator strip tests will benefit from new readout
electronics cards optimised for SiPMs that will allow verification of the MINOS
determined attenuation length and the measurement of single ended readout
efficiency for strips up to 6\u{m} to qualify SiPMs in test beam conditions.
Northern Illinois University has made recent test beam measurements of PPDs
directly coupled to scintillator.  This technology, if successful, would make
the tail-catcher/muon tracker (TCMT) a realistic possibility.  %

The requirements on the beam test setup for prototypes of the muon system are
light, with limited place and narrow beam of mip (up to now a well defined
beam spot (1\u{cm}) of 120 GeV protons at $10^2$--$10^4\u{p/sec}$ is enough).
These conditions are perfectly met at the MTBF at FNAL. There, the set-up can
either be easily shared other R\&D test setup or serve as a complementary
device for e.g.\ a calorimeter prototype such as the CALICE ones.

In the longer term for the ILC there will be more development of readout
electronics (ASIC), SiPM, measurements to improve optical coupling of fibres to
detectors, systems tests with improved DAQ and analysis software and the
evolution of tests with calorimetry and tracking detectors.
And there is the hope that there will be improved tracking at MTBF to better
define beam parameters for individual particles.

\subsubsection{DREAM and dual readout calorimetry}
\label{sec:DualRO}

The DREAM collaboration has tested dual-readout calorimeters in the H4 beam
(North Area) at CERN from 2004 through 2009~\cite{dream}.  %
These tests started with the small 1\u{kt} DREAM module (consisting of Cu tubes
filled with scintillating and clear fibres), and resulted in publications on
basic responses and resolutions~\cite{dream}(a-c), shower
shapes~\cite{dream}(d,h), scintillation-\v{C}erenkov separation in
fibres~\cite{dream}(e), and the response to and role of neutrons in a
dual-readout fibre calorimeter~\cite{dream}(f,l).

Further test included dual-readout in crystals: Lead Tungstate (PWO, or
PbWO$_4$) single crystals~\cite{dream}(g,i,j,k,n,p), extending to arrays of
PWO and BGO crystals~\cite{dream}(m,o), and finally including a full mock-up of
a crystal-plus-fibre calorimeter with $11\u{\lambda_{int}}$
depth~\cite{dream}(o).

The DREAM collaboration has continued testing in the H8 beam at CERN in July
2010 and will continue for 3-to-5 years exploring the ultimate hadronic energy
resolution attainable in dual-readout calorimeters.

The measurements taken in the one-week H8 test run in July 2010 included
(a)~direct comparisons of BGO and BSO crystals; (b)~measurements of the
response variations among eight doped PWO crystals of nominal identical
manufacture; (c)~tests of an ``anti-\v{C}erenkov'' PMT; (d)~tests of a
Pb-quartz plate module; and, (e) direct measurement of polarised \v{C}erenkov
light in a BSO crystal.  These studies will be published soon.

Finally, a large dual-readout fibre module with an expected 1\u{\%} average
leakage is being built to complement and complete the measurements made with
the small 1\u{kt} DREAM module.  We also expect to build a crystal ``em''
module to test in conjunction with the larger fibre module.  %
These activities are scheduled for the H8 beam at CERN.

\subsubsection{Forward calorimetry}
\label{sec:FCAL}


The FCAL collaboration~\cite{FCAL} develops technologies for the
instrumentation of the very forward region of detectors at the ILC or CLIC
collider.  %
For the validated detector concepts ILD~\cite{ild-loi} and SiD~\cite{sid-loi}
two calorimeters are foreseen: LumiCal for a precise luminosity measurement and
BeamCal for a bunch-by-bunch luminosity and beam-parameter estimate.  %
For the latter the depositions from beamstrahlung pairs in BeamCal are used.
For the measurement of the beamstrahlung pair density BeamCal will be
supplemented by a pair monitor consisting of a layer of pixel sensors in front
of BeamCal.

Both calorimeters extend the detector coverage to low polar angles, potentially
important for search experiments using missing momentum as signature.  The
challenges are high precision shower position measurement in LumiCal, radiation
hard sensors for BeamCal, and a fast front-end electronics for both.
  
\begin{itemize}
\item The {\em LumiCal} dedicated to the precise measurement of the luminosity using
  Bhabha events.  It features 30 tungsten layers of 1\u{X_0} thickness each,
  interspersed with very finely segmented silicon sensors.  %
  To ensure a precision of the luminosity measurement of $10^{-3}$, as required
  from physics, a correspondingly precise shower position measurement is
  needed.  The latter can be translated in severe constraints on the sensor
  positioning accuracy and the position monitoring of the calorimeters.  Due to
  the relatively high occupancy fast front-end electronics and digitisation is
  needed;

\item The mechanical structure of the {\em BeamCal and Pair Monitor} is similar to
  LumiCal.  However, due to the large depositions from beamstrahlung pairs,
  about $10\u{Mgy/yr}$ for the sensors near the beam-pipe, radiation
  hard sensors are needed. For this purpose large area GaAs sensors are under
  development in collaboration with partners in Russia.  Also CVD diamond
  sensors are investigated.  The BeamCal has to be readout after each bunch
  crossing.  In addition a fast signal added up from groups of pads is foreseen
  for beam-tuning. Specialised fast front-end electronics is under development.

  The Pair Monitor is a pixel sensor covering the front area of BeamCal. %
  SoI technology is chosen with readout integrated in the silicon wafer;

\item The system may be completed by a {\em GamCal}, 100\u{m} downstream of the
  detector, is considered to assist beam-tuning.
\end{itemize}

To investigate the radiation hardness of several sensor materials a special
test-beam program is ongoing.

Major components of BeamCal and LumiCal, as sensors, flexible PCB for signal
transport, front-end ASICs and ADC ASICs are available as prototypes and tested
separately.
Just now a full system comprised by sensor prototypes and an acquisition
chain is being mounted. First measurements in the 5 GeV electron test-beam
at DESY are foreseen in August 2010, using the EUDET telescope.

\noindent
Test-beam requirements:
\begin{itemize}
\item For irradiation studies electron beams with currents between 10 and
  100\u{nA} and around 10--40\u{MeV} energy are appropriate.  %
  Such beams are available, and used, at the sDALINAC at the Technical
  University in Darmstadt, and at the ELBE linac at Forschungszentrum
  Dresden-Rossendorf (FZD);
\item For performance studies of fully assembled sensor planes the 4\u{GeV}
  electron beam with a beam intensity of a few 10\u{s^{-1}} seems sufficient.
  Such a beam is available at DESY.  In 2010 two weeks are scheduled. Similar
  campaigns are planned in the following 3 years;
\item Within AIDA the plan is to prepare a prototype of a calorimeter sector.
  To test its performance a electron beam with energies comparable with Linear
  Collider beam energies, as available e.g.\ at CERN, will be needed.  This
  program is foreseen to start in 2012.
\end{itemize}

\subsubsection{Summary on tentative sites and special requirements}

The beam test campaigns for the CALICE physics prototypes will be conducted
initially at FNAL in 2010 with physics prototypes and continued at CERN in 2011
with technological ones.  %
The natural preferred site for the beam tests to be conducted with the
technological prototypes is CERN since most of the R\&D groups involved in
these prototypes are based in Europe.  %
As it is currently however difficult to predict fully the availability of the
CERN facilities, FNAL remains a serious option for a test beam site.  %
The prototypes of the CALICE collaboration will not need a dedicated ILC like
beam structure.  Rather it is desirable to obtain beams with a relatively long
flat top with an intensity of not much more than 2\u{kHz}.  Such a
configuration would is demanded by the layout of the front end electronics
which is designed for low occupancy.  %
The validation of the power pulsing technique will however need the
availability of a large bore magnet ($\diameter>1\u{m}$) with a field strength
between 3 and 6\u{T}.  %
In addition beam telescopes with an excellent point resolution should be part
of the beam line equipment. \\
The program for muon detectors in the coming years will be based at FNAL and
has no special requirements on beam conditions.

Test beams with the prototype of the SiD Ecal will initially be conducted at
SLAC with the option to move to FNAL for beam tests with hadrons. The design of
the front end electronics for this prototype renders highly desirable the
availability of an ILC like beam structure.
Low rate beams are mandatory. \\
Test beams with the dual readout technique will be continued at CERN by the
DREAM collaboration in the coming 2--3 years.  %
Here, hadron beams up to the highest energies will be needed.\\
In the coming years the forward calorimeters will sustain irradiation tests
with low energy but high intensity electrons beams or electrons beams in the
few GeV range.  These beams are available at Darmstadt, Dresden and DESY.  %
The detector response will be tested with low intensity electrons first at DESY
at low energy then at CERN with high energy.\\
All plans of calorimetric and muon systems described before are summarised in Table~\ref{tab:calmuplans}.
\begin{landscape}
  \begin{table}[htdp]
    \begin{tabular}{@{} |l|l|l|l|l|l| @{}}
      \hline
      Calorimeter 		   & Date 			 & Type	 		   & Requirements	                & Projected TB facility (optional) \\ 
      \hline
      RPC DHCAL m$^3$ ($\varphi $) & ${\geq}$ mid 2010 	         & All types 		   & \multirow{2}*{$<100\u{Hz}$}	& \multirow{2}*{FNAL}              \\ 
      ~                            &                             & High E (in combined TB) &                                    &                                  \\
      \hline
      GEM DHCAL ($\varphi $) 	   & ${\geq}$ 2011 	         & low E e, $\mu $, $\pi $ & {}---		                & FNAL	                           \\ 
      \hline
      {$\umu$}Megas, RPC layers    & 2009 $\rightarrow$ end 2010 & low E e, $\mu$, $\pi$   &                                    & CERN 	                           \\ 
      SDHCAL m$^3$ ($\tau $) 	   & ${\geq}$ end 2010           & All types 	           & $<100\u{Hz}$ or ILC like           & CERN (FNAL)	                   \\ 
      \hline
      W HCAL structure ($\varphi$) & ${\geq}$ '10 		 & All types               & {}---   		                & CERN	                           \\ 
      \hline
      DECAL  			   & ${\geq}$ 2011               & e (all E)  		   & Large XY table	                & CERN \& DESY                     \\ 
      \hline
      CALICE AHCAL ($\tau$) 	   & ${\geq}$ 2012		 & e (all E), low E $\pi $ & $\leq1\u{kHz}$ or ILC like         & CERN (FNAL)                      \\ 
      \hline
      CALICE ECALs ($\tau$) 	   & ${\geq}$ 2011		 & e (all E), low E $\pi $ & $\leq1\u{kHz}$ or ILC like         & CERN (FNAL)                      \\ 
      \hline
      Combined CALICE ($\tau$) 	   & ${\geq}$ 2011-2012 	 & All types 		   & $\leq0.1-1\u{kHz}$ or ILC like     & CERN \& FNAL                      \\
      ~ 			   &                             &                         & $>3\u{T}$ magnet, telescope        &                                  \\ 
      \hline
      \multirow{2}*{SiD ECAL}      & ${\geq}$ 2011 	         & e 5--10+ GeV 	   & Beam localisation 	                & SLAC (DESY)                      \\ 
      ~				   &                             & low E e, $\pi$ (FNAL)   & ILC like, low rate (0,1,2 e/Bunch) & FNAL                             \\ 
      \hline
      SiD Muons 		   & ${\geq}$ 2011		 & High E had.		   & {}--- 		                & FNAL	                           \\ 
      ~				   &                             & Combined test 	   &                                    & FNAL	                           \\ 
      \hline
      FCAL 			   & 2010--2013 		 & low E e		   & Telescope 		                & DESY                             \\
      ~				   & $\ge$ 2012		         & High E electrons	   & Telescope		                & CERN                             \\ 
      ~                      	   &                             & Irradiation with e	   &                                    & FZD, TU Darmstadt                \\
      \hline
      DREAM            		   & 2010--2013                  & High E had. 	   	   & {}---         	                & CERN 	                           \\
      \hline
    \end{tabular}

    \caption{\sl Prototypes ($\varphi$ and $\tau$ refer respectively to Physics and
      Technological CALICE prototypes), date of first test beam operations, run
      types \& constrains, estimated time.}
   \label{tab:calmuplans}
  \end{table}

\end{landscape}



\subsection{Silicon tracking}
Silicon-based tracking and vertexing is continuing to develop over a 
broad front.  Silicon tracking detectors are well-placed to take advantage of 
rapid development in silicon technology.  These new technologies need to 
be developed, tested, and validated in test beams.  Some technologies, 
like the DEPFET have already demonstrated resolution less than 5 microns 
and require high momentum beams and sophisticated telescopes to 
make proper measurements.  In parallel tracking detectors are testing 
larger and more realistic "ladder" designs and will need realistic infrastructure 
such as pulsed power, ILC-like beam structure, magnetic field, and low 
mass supports.

There is a broad range of work on vertex and tracking technology.  Table~\ref{tab:sitr} summarizes some of the technologies being studied for the ILC tracker and vertex detector.\\
\begin{table}
\begin{tabular}{| p{4 cm} | p{5cm} | p{3.75cm} | p{1cm} |}
\hline
Group        & Technology                          & Goals                & Test Beam  \\
\hline
SiD tracking & Multi-metal strips + KPIX chip      & SID Outer Tracker    & FNAL       \\
DEPFET       & Depletion mode FET                  & Belle-II, ILC Vertex & CERN       \\
MIMOSA       & CMOS MAPS development               & ILC Vertex           & DESY, CERN \\
SPYDR        & CMOS MAPS, deep n-well              & Tracking and Vertex  & ?          \\
3D           & 3D detector/electronics integration & ILC Vertex           & FNAL       \\
APSEL        & CMOS MAPS triple well, 3D           & ILC Vertex           & CERN       \\
CAPS         & CMOS MAPS + SOI                     & ILC Vertex, Belle 2  & FNAL       \\
Thinned MAPS & CMOS MAPs thinning                  & ILC Vertex, RHIC     & FNAL       \\
SiLC         & Silicon strips                      & ILC (ILD) Tracking   & CERN       \\
FPCCD        & Fine pixel CCD                      & ILC Vertex           & KEK?       \\
ISIS (LCFI)  & CCD with in-pixel storage           & ILC Vertex           & ?          \\
CPCCD (LCFI) & Column-parallel CCD                 & ILC Vertex           & ?          \\
Chronopixel  & CMOS MAPS                           & SID Vertex           & ?          \\

\hline
\end{tabular}\\
\caption{\sl Overview on the projects and beam test plans of the various groups working on Silicon Tracking and Vertex Detection. The ILC tracking and vertex detector reviews include a more comprehensive review of the efforts of the different R\&D groups.}
\label{tab:sitr}
\end{table}%

\subsubsection{Beam properties and structure}
The ILC has a very distinct time structure, with a train of 2820 bunches separated by 
337\,ns followed by a $\approx$ 200\u{ms} gap.  Such a structure is difficult, but not impossible 
to mimic in a test beam.  Depending on the application, the ILC structure could be 
mimicked by appropriate trigger electronics or offline analysis. How well this works 
depends on the  details of the detector integration time, time stamping ability, and 
saturation effects.  Many aspects of pulsed powering could be tested independent of 
beam conditions. History has shown that detailed tests in an environment as close 
as possible to actual operation are invaluable.

Other beam properties are also important.  High energy beams are the only way to 
unambiguously test detector resolution with minimal multiple scattering.  However 
lower energies are also important to quantify the scattering and validate Monte 
Carlo models of the detector response.  Beams should be able to simulate the rates 
seen at the inner radius of the vertex detector.  

Two-track resolution needs to be studied, both for normal and for glancing incidence. 
This can be done in a high rate beam, using multiple tracks which pass through the 
detector within the integration time, or by a secondary target which mimics the interaction 
vertex.  In the case of a secondary target all relevant tracks in the event need to be 
reconstructed.  The momenta also probably have to be measured.  This makes for a 
much more complex setup with a significant magnetic field.

\subsubsection{Beam instrumentation}
A high quality beam telescope is needed to determine the reference position of the charge particle track. For an unambiguous measurement of the spatial resolution of the device this position must be precisely predicted. Especially for state-of-the-art vertex detector technology this latter requirement poses a severe challenge, requiring sub-micron pointing precision. 

Traditionally, much effort of the R\&D collaborations is devoted to the construction of precise beam telescopes. The EUDET project~\cite{eudet} offered a precise telescope~\cite{Roloff:2009zza} based on MIMOSA monolithic active pixel detectors, hardware to synchronize devices under test and the telescope to the trigger signal as well as a flexible DAQ environment.  This infrastructure has attracted a large user community~\cite{Gregor:2009yy}. 

In AIDA~\cite{aida} this common infrastructure will be continued and extended. A flexible telescope, combining the precise and thin MIMOSA sensors with fast ATLAS hybrid pixel detectors and/or time-stamping TimePix devices will be built. A $CO_2$ cooling system will be provided for the test of large-scale prototypes and irradiated devices. 

Future applications require devices that combine performance (resolution, read-out speed) with an extreme control of the material in the tracking volume. As more transparent devices are developed the mechanical and thermal design becomes more and more challenging. To characterize the thermo-mechanical properties of prototypes under realistic powering and cooling conditions, a second infrastructure will be developed in AIDA. 

Finally, it is envisaged to install silicon $\mu$-strip detectors in front of the highly granular calorimeter infrastructure, described in Section~\ref{calo}. These layers aim to provide a precise entry point, thus aiding the analysis of overlapping showers.

A flexible readout system will also be important for testing the large variety of devices being brought to the beams.  One example is the CAPTAN system~\cite{CAPTAN, Turqueti:2008zz, Rivera:2009zz}, developed by FERMILAB, and designed a a flexible, FPGA-based readout system for a variety of devices.  To date the CAPTAN has been used to read out the 
BTeV FPIX chip, the CMS pixel chip, the VIKING strip chip, and will be used for 
the VICTR CMS track trigger chip. 

The ALIBAVA system~\cite{alibava} provides a flexible read-out system for $\mu$-strip detectors. Originally developed for the read-out of (irradiated) test structures and small-scale devices (so-called {\em baby} detectors), the system is being upgraded to allow multi-module operation with an external trigger, suitable for test beams. A prototype sensor is wire-bonded to a board that includes a Front-End chip (the Beetle of the LHCb VELO detector) and all the ancillary electronics to read-out and control the Front-End. The system is controlled from a PC through a standard USB connection.

 Small area trigger can be provided by a version of the VICTR chip, which has a 
 array of 64 $1\u{mm}\times100\u{\umu m}$ strips with a fast output and maskable pixels.  The chip is designed (for LHC upgrade triggering) for a coincidence with a second detector 1-2\,mm away.
 
 


\subsection{Gaseous Tracking (2 pages - T. Matsuda)}
\label{sec:gastr}
Physics at the International Linear Collider (ILC) or the Compact Linear Collider (CLIC) will require 
a detector of high precision. A tracking system of the detector has to achieve a high momentum 
resolution $\delta(1/p_{t})$ of a few $10^{-5}$\,$\rm{(GeV/c)^{-1}}$~\cite{tpc-res}. 
This resolution surpasses by 10 times the best momentum resolution achieved by the experiments at LEP. 
The tracking system should also provide a high tracking efficiency down to a few GeV/c to ensure a good jet-energy 
measurement by the Particle Flow Algorithm (PFA) in an environment of high beam-induced backgrounds.

To meet with these requirements, a large Time Projection Chamber (TPC) with using  Micro Pattern Gas Detectors (MPGD) 
is proposed as a central tracker of the International Large Detector (ILD)~\cite{ild-loi}. The ILD TPC is to be located in a large 
superconducting solenoid of 3.5\u{T}. It measures each track at 220 space points with an $r\phi$ spatial resolution of 
100\,${\rm \mu m}$ or better in the whole drift volume of 2.2\,m length. This performance of TPC is only achievable with the MPGD 
technology~\cite{mpgd-1, mpgd-2}. At this moment three candidates of MPGD detectors are considered; Bulk MicroMEGAS with resistive anode 
readout, GEM with narrow pad readout, and, in a somewhat longer time scale, a digital TPC with Ingrid TimePix or a semi-digital 
TPC using GEM readout by TimePix. 

\subsubsection{TPC R\&D by the LC TPC collaboration}

The LC TPC collaboration has been carrying out  R\&D of the MPGD TPC for ILC (ILD) in three stages; 
\begin{enumerate}
\item Demonstration phase. 
\item Consolidation phase. 
\item Design phase. 
\end{enumerate}
At each phase for the last several years, a multitude of beam tests have or will be conducted. 

In the demonstration phase (2004-2007) a basic evaluation of the properties of the MPGD gas amplification 
was made, demonstrating that the requirements for the linear collider (ILD) could be met. For an example, 
it was shown through beam tests with small TPC prototypes that the $r\phi$ space resolution of 100$\umu$ could be
achieved by both, MicroMEGAS with the resistive anode readout and GEM with the narrow pad readout~\cite{npro-1, npro-2}.

In the current consolidation phase (2007- ), a TPC Large Prototype 1 has been successfully operated (TPC LP1) 
at the low energy electron 5\u{GeV/c} test beam at DESY, T24-1~\cite{klaus, delent}.  The goals of the LP1 beam test are to 
confirm the results from the demonstration phase for a larger scale TPC, and to show that the excellent momentum resolution 
is actually achievable for the LC TPC. In 2010 beam tests are performed with the LP1 end-plate equipped with four 
to seven MPGD modules, and a new TPC tracking code for non-uniform magnetic field is under development. In this phase, 
in addition to the development of the different MPGD TPC readout modules, basic engineering issues for the 
LC TPC are studied. Good examples are the construction of a thin LP1 field cage and the development of a low noise, high-density TPC 
pad readout electronics using S-ALTRO~\cite{saltro}. 

Currently the collaboration is entering the design phase (2010- ) during which a basic conceptual design of the LC TPC is worked out.  

\subsubsection{LC TPC R\&D and Beam Tests in 2010-2012}

In the design phase of 2010-2012, beside the overall design of the LC TPC features two major hardware R\&D issues; (a) a design of a TPC 
end-plate of the thickness of 15\% radiation length or less, and (b) a choice of the ion gating device. A study of a 
light mechanical structure of the TPC end-plates has been started as well as the so-called advanced end-plate TPC modules with power pulsing and an 
efficient cooling such as the two-phase CO2 cooling~\cite{co2cool}. The plan is to build a new LP1 end-plate structure mounted with 
the advanced TPC  modules with S-ALTRO (and also modules of the digital TPC), and test it in a beam test for the ILD-DBD (Detail Baseline Design) 
in 2012. This R\&D phase is not yet a fully funded. In 2011, the magnet PCMAG will be modified in order to make it a 
superconducting magnet without liquid He supply.  The modification will take about 6 months.

\subsubsection{Test beam before 2012 and the ILC beam structure}

In the current prospect of the R\&D budget and support, and in the situation where the availability of a higher energy hadron 
test beam in 2011 seems to be not very clear, it is planned that the TPC LP with PCMAG stays at the T24-1 beam line at DESY until 
to the end of 2012. Optionally, small scale beam tests at high-energy hadron test beams and in a higher magnetic 
field using small prototypes will be organised on short notice.

At the DESY test beam, there is no plan to simulate the ILC beam bunch structure. The power pulsing of the advanced 
TPC modules will be tested with the beam without the ILC bunch structure. The opinion of the collaboration is that a beam to test the power pulsing is not really needed. 
The functional test of the power switching of the advanced TPC modules in a higher magnetic field will be necessary. To demonstrate 
how the ion gating device works, a proper device is required, either a laser or a flash lamp, to simulate beam backgrounds at ILC 
according to the beam bunch structure.

\subsection{Data acquisition}
\label{sec:daq}

In general, for a given ILC sub-detector, a dedicated data acquisition (DAQ) system is 
developed to suit its needs, depending on a multitude of technical issues such 
as data rates, number of channels to read out, etc.  The DAQ system consists of 
the hardware---various electronics boards using various standards to get the data 
from the detector head to a PC---and software to control the flow of data from 
the detector and commands to the detector.  The requirements can then lead to 
a DAQ system which is conceptually new or is strongly based on an existing system 
in use for another detector; both of which are reasonable approaches.  This 
therefore results in very different systems when developed in isolation as is 
the case for several of the ILC sub-detectors; a brief review of some of the 
systems is given below.  

Were sub-detectors to continue in isolation a programme of verification in a 
beam-test, then bespoke development is a sensible approach.  
However, should any sub-detectors wish to have combined beam-tests with another 
sub-detector, then more thought and planning is needed.  Therefore any issues 
with regard to DAQ systems depend crucially on whether combined beam-tests of 
several sub-detectors will happen.  Alternatively, given extra resources such as 
those provided by the AIDA project, a common approach to DAQ systems 
can be pursued now such that a final system for a final ILC detector will be 
easier to manage and integrate when it becomes a reality.  Careful planning now 
could lead to significant benefits, with reduced risk, in the future.  As a DAQ 
system serves a given sub-detector or detector, it is not a driver for individual 
or common beam tests which is dictated by the detectors themselves.  As a 
separate goal, more generic aspects of DAQ system can be developed for future 
sub-detector use which will save on effort in the long-run.

\subsubsection{Example DAQ systems}

\subsubsection*{\underline{CALICE DAQ}}

Most of the focus for the new CALICE DAQ system has been on the hardware development 
and firmware to control it~\cite{wing-memo,decotigny-lctw09}.  The system consists 
of several layers of concentrator 
cards to get the data from the detector head to a PC and storage.  Given that the 
CALICE programme includes several different types of calorimeter, the first layer 
of electronics needs to convert the sub-detector-specific data into a generic 
structure which is then passed to the next layer.  As such, the hardware system 
needs to be suitably generic and could in principle be used for various 
sub-detectors and not just calorimeters.  The DAQ and slow control software are 
less advanced.  Initially the approach was to use existing software designed to 
cope with large-scale systems; the programmes DOOCS~\cite{doocs} and 
XDAQ~\cite{xdaq} have been used so far.  In light of possible combined beam-tests, 
a survey of available software is being performed.

\subsubsection*{\underline{EUDAQ system for vertex and tracking detectors}}

A DAQ system developed to read out the EUDET~\cite{eudet} pixel telescope has been 
developed~\cite{haas-report,haas-talk}.  The telescope is a relatively small-scale 
detector and is read out via a VME-based hardware system.  Major effort has been 
invested in writing a flexible DAQ software framework, called EUDAQ, which has been 
successfully used for the pixel telescope in numerous beam tests.  The code is 
written in {\tt C++}, is freely available and was fully developed by the main 
authors.  The software has 
been used by several other groups when performing beam tests in conjunction with the 
pixel telescope.  Indeed the LC-TPC collaboration are using it for their work on 
a TPC sub-detector~\cite{killenberg-talk}.  Any new sub-detector just needs to 
write a producer and the EUDAQ authors should be able to integrate on the time-scale 
of a few days.

\subsubsection{Towards a common DAQ system}

As sub-detectors will at some point be used together, say as a complete detector 
slice-test, the data will have to be merged at some point.  The extremes are : to 
develop one data acquisition system, both hardware and software, which is able to 
read out all sub-detectors; or for sub-detector DAQ systems to all be developed in 
parallel and data merged at the final opportunity when it is stored.  The former 
is unlikely given the various logistical problems whilst the latter is undesirable, 
potentially leading to wasted effort and a lack of coherency in the final data 
samples.  The reality will lie somewhere in between with some common hardware used 
and even more so, common software.  From the examples given above, the CALICE 
hardware could in principle be used for other sub-detectors, although this would have 
significant costs associated to it.  The EUDAQ software may be a viable solution for 
CALICE calorimeters, although this needs to be demonstrated given its current use for 
a much smaller system.  As DAQ systems for all sub-detectors are relatively well 
advanced, adapting to common solutions will require extra effort and will require 
e.g. the recent funding of AIDA to make it possible.

Taking a middle ground on common aspects of a DAQ system, some of the questions and 
issues which need to be addressed are listed below.  These should be addressed in the 
AIDA project.

\subsubsection*{\underline{Common hardware}}

Although the hardware used for the CALICE calorimeters and the 
CAPTAN~\cite{CAPTAN} project are relatively generic and 
could be used for other sub-detectors, it is unlikely that such an approach is 
possible.  However, there are various common items amongst the various sub-detector 
groups which could be used :

\begin{itemize}

\item Hardware which provides a trigger or a clock such as the Trigger Logic 
      Unit~\cite{haas-report} or Clock and Control Card~\cite{wing-memo} 
      developed for the pixel telescope and CALICE calorimeters, respectively, 
      could be used by all sub-detectors.  These would uniquely identify each 
      trigger.

\item A proposed ``Beam Interface Card''~\cite{decotigny-lctw09} could be used 
      to monitor beam conditions taking data from e.g. scintillators, hodoscopes, 
      etc.  Its exact form is to be designed.

\end{itemize}

\subsubsection*{\underline{Common software}}

There are a multitude of DAQ software frameworks developed for previous or 
existing experiments.  A critical review of these needs to be done :

\begin{itemize}

\item Large software frameworks such as XDAQ, DOOCS, TANGO~\cite{tango}, etc. 
      have been developed with large-scale, diverse apparatus in mind.  
      Presumably they then have the necessary functionality and flexibility to 
      provide the framework for the ILC sub-detectors.  This needs investigation 
      and the various software compared;

\item The EUDAQ software has been shown to work successfully with a number of 
      different sub-detectors.  However, its efficacy for reading out large 
      systems such as the CALICE calorimeters, with thousands of channels, 
      must be verified;

\item Information needed to decide on the nature of the read-out path is 
      the data volume, zero suppression, compression, data format etc.;

\item It is generally agreed that all data should be converted into the common ILC 
      offline software format, currently LCIO.

\end{itemize}

In summary, commonality between the DAQ systems of the various sub-detectors should 
be sought at an early stage so as to ease integration later.  Given the funding of the 
AIDA project, this will give support to this effort in which a critical 
review of current DAQ hardware and software is carried out leading to a more coherent 
framework for future ILC detector beam tests.



\subsection{Software}
Software development for ILC test beam experiments has a
large potential for collaboration, as typical computing tasks in high
energy physics event data processing have a high degree of similarity
from experiment to experiment. For example every experiment needs a
way to store and retrieve the conditions data, defining the
experimental setup at the time of data taking.
In order to avoid duplication of effort, most of the current test beam
collaborations are already using a common set of core software tools.
This desirable development has been greatly fostered by the EUDET~\cite{eudet}
project during which already existing software tools have been
improved and combined into a common framework, referred to as ILCSoft~\cite{marlin_et_al}. 
The same software framework is also used by the ILD detector concept,
the CLIC detector working group and in parts by the SID detector
concept. These groups work on the development and optimisation of the
global detector concepts, based on Monte Carlo simulations and results
from the R\&D test beams. Having a joint software framework thus
provides synergies for both communities, as code and knowledge can be
shared easily and provide for the necessary feedback of realism into
the full simulation.\\ 

\subsubsection{ILCSoft tools}
\begin{figure}[ht]
\begin{center}
\includegraphics[width=0.5\textwidth]{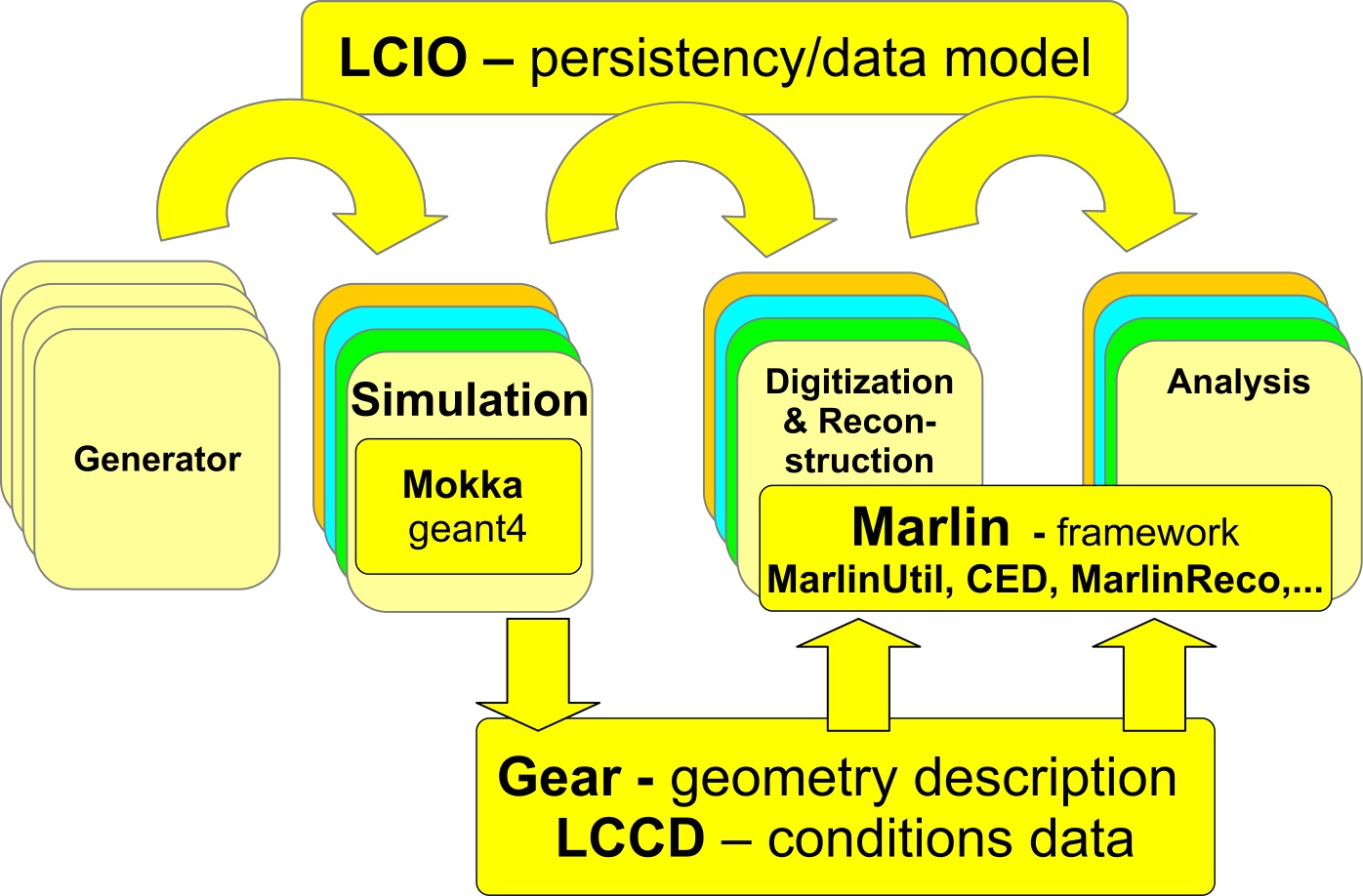}
\caption{\sl Schematic overview of the ILCSoft framework tools.}\label{fig:ilcsoft}
\end{center}
\end{figure}

ILCSoft is based on LCIO~\cite{lcio}, which is a persistency file
format for ILC studies and defines a hierarchical event data model for
full detector simulation and dedicated raw data classes for beam test experiments.
The core of the ILCSoft framework is defined by Marlin, a modular
C++ application framework that uses LCIO as its transient and
persistent event data model. 
Marlin is complemented by a number of software tools: GEAR which
provides the high level view other detector geometry and materials as 
needed during reconstruction and analysis, LCCD a conditions data
toolkit that provides access to the conditions data
and CED a fast 3D event display.
The simulation of the detector response is performed in the
GEANT4 application Mokka~\cite{mokka}. The geometry description in Mokka is interfaced to GEAR
for reconstruction and analysis.
Fig.~\ref{fig:ilcsoft} shows an overview of the main tools used in ILCSoft.
The core framework is completed through a number of auxiliary tools, such as RAIDA for
histograming and the utility package MarlinUtil and depends on a small set of external packages
like ROOT, GSL and CLHEP.\\
The following planned developments and improvements for LCIO are currently ongoing:
\begin{itemize}
\item{Direct access to events;}
\item{Splitting of events and partial reading of event data;}
\item{Streaming of user defined classes which is particular useful during the development phase of a detector.}
\end{itemize}
Using ROOT I/O for the implementation of these new features is under investigation.
Another area of possible improvement is the geometrical description of the
detector. While the current system ensures one leading source of the
geometry, the Mokka simulation, it could be made more flexible by
having a standalone tool that feeds into simulation, reconstruction
and event displays. The development of such a flexible system is
foreseen in the proposed AIDA project. This would also include
mis-alignment and integration with conditions data as the distinction
between geometry and conditions data is not always perfectly well defined.

\subsubsection{CALICE and LC-TPC software}
The CALICE collaboration was the first test beam group to adopt
the ILCSoft framework. CALICE has been using the complete framework for
their past data taking campaigns and provided very useful feedback
that led to the improvement of the software tools in particular in the
context of the EUDET project. CALICE is not using LCIO as their raw
data format, but are converting their data to LCIO within hours of the
data acquisition. This 'duplication' of raw data has proved to be less than
optimal and having one raw data format only would be desirable for
future beam tests~\cite{calicesw}. \\
Also LC-TPC was an early user of the common core software tools. They
are currently working on completion of their reconstruction and analysis package
MarlinTPC~\cite{marlintpc}. In that process they improved the geometry
description of the TPC in GEAR in order to meet the requirements. An
example for the fruitful interplay between core software group and
users. LC-TPC also suggested improvements for LCCD, namely to store
the conditions data in data base tables, that can be queried using
MySQL tools.

\subsubsection{Grid computing}
Large computing resources for high energy physics data processing will
be available only on the Grid. All the test beam data that has been
accumulated so far is stored on Grid storage elements and major Grid
sites did provide so far sufficient computing resources for their
analysis. This was partly facilitated due to the delay of the LHC, for
which massive resources had been allocated. With the LHC now running 
it is important to make the Grid sites aware of the computing needs of 
upcoming ILC beam tests so that they can plan accordingly.\\
\subsubsection{Remote control and communication tools}
Besides data analysis software for beam tests, control and
communication tools are an important aspect that can foster
collaboration and reduce travel expenses. A nice example is the CALICE
control room that was recently set up at DESY~\cite{calice_control}
and is fully functional from the start. This room was realised for
comparatively small budget, that paid off in a short period of time
through savings in travel cost. \\
With improvements in audio and video technologies, increased band
widths and lower cost, modern communication tools and remote control
centres will become more widespread and are likely to change the way
experiments are run.

\section{Sites}
\subsection{CERN}
CERN offers a broad range of test beam facilities with beams originating both from the Proton Synchrotron (PS) 
and Super Proton Synchrotron (SPS) accelerators. At the CERN PS East Hall, there are two beam lines, T9 and T10, 
delivering hadrons, electrons and muons of up to 15\u{GeV/c} and 7\u{GeV/c} momentum respectively. During a spill length of 
400 ms, occurring typically every 33\u{s}, up to 106 particles can be delivered. Recent studies have indicated that an 
ILC-like beam structure can be produced at the PS. In the SPS North Area hall EHN1 there are four beam lines 
(H2, H4, H6, H8) with several experimental areas each. The H2, H4 and H8 lines can provide secondary hadrons, 
electrons or muons of up to 400\u{GeV/c} or primary protons of up to 450\u{GeV/c}. The H6 line has a maximum momentum of 
205\u{GeV/c}. Up to $2\times10^8$ particles per spill can be delivered. Spill lengths vary from 4.8 to 9.6\u{s}, while spills 
are repeated every 14 to 48\u{s}, depending on the number of SPS users. 
Together with the beams themselves, CERN provides some adjacent infrastructures, such as basic beam instrumentation. 
These comprise beam spectrometers for precise momentum definition, wire chambers to measure beam profiles, as well as 
threshold \v{C}erenkov counters and Cedar counters for particle ID. On request a scanning table can be provided and some 
beam lines are equipped with magnets, which can surround the equipment under test.
In 2010 the PS and SPS are scheduled to provide 28 weeks of beam. Since many years, the CERN test beams have been used 
extensively by the linear collider detector community. This tradition continues. In 2010 a total of 28 days are scheduled 
for linear collider-related tests at the PS T9 beam, 34 days at the SPS H4 beam and 48 days at the SPS H6 beam. 
The linear collider users represent several CALICE HCAL technology tests, SiLC tests and various vertex technology tests. 
For the following years, the PS and SPS test beam schedules are expected to have some dependency on the LHC schedule, 
with most likely a similar availability of test beams in 2011 and potentially a somewhat shorter duration in 2012.  
Users have two ways to apply for beam time. For short beam tests, $<2$ weeks at the PS or $<1$ week at the SPS, requests 
are addressed directly to the SP/SPS coordinator (sps.coordinator@cern.ch) by submitting a form. These requests are 
normally collected towards the end of the year for the following year. For beam tests of longer duration a formal 
request has to be addressed to the SPSC committee.
Some user groups have semi-permanent beam test installations. Examples are the CMS experiment in the H2 line, the 
ATLAS experiment in the H8 line and the RD51 collaboration in the H4 line. Following approval by the SPSC, these 
installations have been built up through a common effort by the collaborations involved. 
What concerns the linear collider activities, the establishment of semi-permanent ILC beam line at CERN, should be requested latest by mid-2010 in order to have it available by middle of 2011.
\subsection{DESY}
DESY provides three electron beam lines with an energy range from 1 to 6\u{GeV}. The beams are produced at the
DESYII synchrotron which mainly serves as injector for the DORIS and PETRA accelerators and has typical up-times
of 10-11 months per year. The high availability and flexible scheduling - related to intensive in-house use - are major
assets of these facilities. The beam is delivered in short 30\u{ps} bunches every 160 or 320\u{ms}, with typical event rates
of 1\u{kHz}. All beam lines are equipped with pre-installed cables, fast networking and installations for pre-mixed gases.
Moving stages, gases magnets and beam telescopes can be provided upon request, while users in general bring their
own DAQ and trigger hardware. In the previous years, the infrastructure has been considerably enhanced in the framework
of the EUDET initiative. The refurbished are T21 hosts the EUDET pixel telescope, while the upgraded ZEUS telescope serves
users in T22. The super-conducting PC magnet provides a field of 1\u{T} in a bore of 0.85\u{m}. With no iron yoke, its thickness
corresponds to 0.2\% $X_0$ only. It is presently installed in T24 and heavily used for TPC R\&D (see Section~\ref{sec:gastr}). Following an exceptionally  extended winter shutdown, the machine is running since march 2010 throughout 2010 and is expected
to have high availability also in the forthcoming years. Users can apply for beam time through the DESY test beam co-ordinators.  
More information is available under \url{testbeam.desy.de}.


\subsection{Further European sites}
The IHEP at Protvino in Russia provides electron beams between 1 and 45\u{GeV} as well as hadron beams in this energy range.
The site is available for two months in winter time. The beam test facility at Dubna, Russia, provides neutron beams with a good yield. It remains to be discussed how these facilities can be incorporated into the beam test program for linear collider detectors. 

Other sites offering beam test facilities in Europe. These are PSI Villingen (CH), GSI Darmstadt (D), the ELSA beam at Bonn (D) as well as the FZD at Dresden-Rossendorf (D). Some of these were used in the past or will be used in upcoming beam test campaigns.

\subsection{FNAL}
\label{fnal}

\begin{figure}[htbp]
\begin{center}
\includegraphics[width=0.99\textwidth]{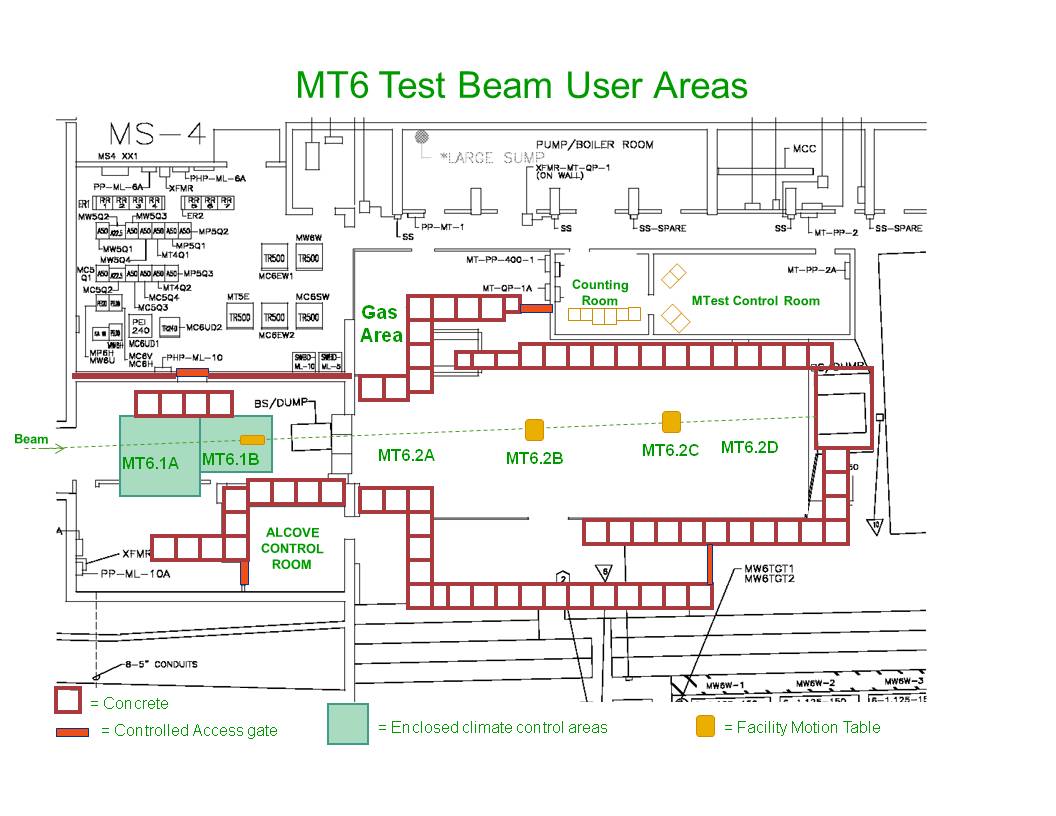}
\caption{\sl Plan view of the Meson Test Facility at FNAL.}
\label{fig:plan_mtbf}
\end{center}
\end{figure}

Crucial to many detector development projects is the ability to test real life operations of the device in a 
high energy particle beam. Only a few such facilities exist in the world.  The United States' only high energy 
detector test beam facility is the one at Fermilab.
The Meson Test Beam Facility (MTest) gives users from around the world an opportunity to test the performance of 
their particle detectors in a variety of particle beams. A plan view of the facility is shown in Figure~\ref{fig:plan_mtbf}.  The web site  for the MTest facility can be found at \url{http://www-ppd.fnal.gov/MTBF-w/}.

\subsubsection{Details of the beam}

The test beam originates from the resonant extraction of at least one Booster batch inside the Main Injector (MI).  
This batch usually consists of 10-60 RF 'buckets', with buckets separated by 19\u{ns}.  Thus the batch is anywhere 
from 0.2-1.2\u{\umu s} long.  The batch is accelerated to 120\u{GeV}, circulates around the MI, and is slowly extracted 
over a macroscopic slow spill using a resonant quadrupole called QXR.  The full circumference of the MI is about 11 microseconds, 
giving a large gap between extractions. 
	The length and duty cycle of the spill is determined by the Accelerator Division (AD), with guidance from the Office of 
Program Planning.  For most operations there is a single 4 second long spill per minute, for a maximum of 14 hours per day.  
The AD has setup a procedure for easily changing from this 4 second spill to a 1 second spill.  This shorter spill can then 
be delivered more frequently for commissioning purposes and for those groups who are data-acquisition buffer limited.  
The AD has also commissioned a "pinged" beam operation where beam is extracted using a pulsed operation of the QXR, 
with up to 4 pings per spill, each with a tunable width from 1 to 5\u{ms}.
	The 120\u{GeV} proton beam has an approximate 0.3\% momentum spread and can be focused to a 7\u{mm} RMS spot size in the 
user area. In addition to delivering primary protons, there are two targets on movable stages that can act as secondary beam 
production areas.  The magnets downstream of those targets can then be tuned to deliver any secondary momentum from 
0.5\u{GeV/c} to 60\u{GeV/c}.  The momentum spread of these secondary beams depends on the energy and the details of the collimation and can range between 1-10\%, 
with the poorer resolution beam occurring for the lower momenta.  The physical size of the beam is approximately 2-5\u{cm} rms for the 
lower momenta.  The Table~\ref{tab:fnalbeam} shows the rate of beam delivered to the user area for some selected momenta.

\begin{table}[htdp]
\begin{center}
\begin{tabular}{@{} |c|c|c|c| @{}}
 \hline
    Beam energy/GeV  & Rate at entrance & Rate at exit & \% $\pi/\mu$ \\
                     & to MT6 (per spill) & to MT6 (per spill) &  at exit of MT6      \\
    \hline
    16& 132000 &  95000  & 82\% \\
    \hline
    8& 89000 &  65000  & 42\% \\
    \hline
    4& 56000 &  31000  & 26\% \\
    \hline
    2& 68000 &  28000  & $< 20\%$ \\
    \hline
    1& 69000 &  21000  & $< 10\%$ \\
    \hline
\end{tabular}
\end{center}
\caption{\sl Rate of beam delivered to the MT6 user facility for $1\times1011$ protons in the 
Main Injector.  Remainder of beam is identified as electrons.}
\label{tab:fnalbeam}
\end{table}%

As part of the improvement in extending momentum range of the beam line, the MINERVA experiment (T977) proposed 
to install an entire new tertiary beam line in the user facility so that it can deliver 300\u{MeV/c} pions onto their test apparatus.  
This beam line was begun in the US FY2008 and has recently been completed. After the completion of the MINERVA 
tests, this beam line will be available for other users.  The target and collimator can be rolled quickly aside 
so that the facility can operate normally.from them as well.

\subsubsection{The future of test beam at Fermilab}

The Meson Test Beam Facility will be in operation for the foreseeable future, since it has demonstrated a wide 
variety of modes of operation. Because the facility is in heavy use, it is likely that additions and upgrades to the 
equipment at MTest will be incremental, with no large update at any given time.
In addition to the Meson Test beam line, Fermilab will be starting a new test beam facility in the Meson Center 
beam line. This facility will be known as the Meson Center Test Facility, or MCenter, and will be used as an adjunct to the 
MTest facility.  The two beam lines are virtually identical, while the user areas are complementary.  While the MTest facility 
has a large variety of user installation areas, and a crane to support them, the MCenter facility is tighter, but has two 
spectrometer magnets that could be used for a variety of calorimetry studies.  Currently the MIPP experiment's apparatus 
occupies the downstream location in MCenter. This apparatus could be used to perform tagged neutron studies, as well as 
support tracking for more advanced installations. With the help of a thin target a ''jetty'' environment could be mimicked 
for future beam tests. Fermilab has begun efforts to provide for a user facility in MCenter to support detector R\&D.  
With a very successful MTest beam line, and a second MCenter beam line to augment it, then Fermilab's test beam facilities 
will remain in the forefront of detector support in the United States for quite some time.

\subsection{SLAC}

\begin{figure}[htbp]
\begin{center}
\includegraphics[width=0.99\textwidth]{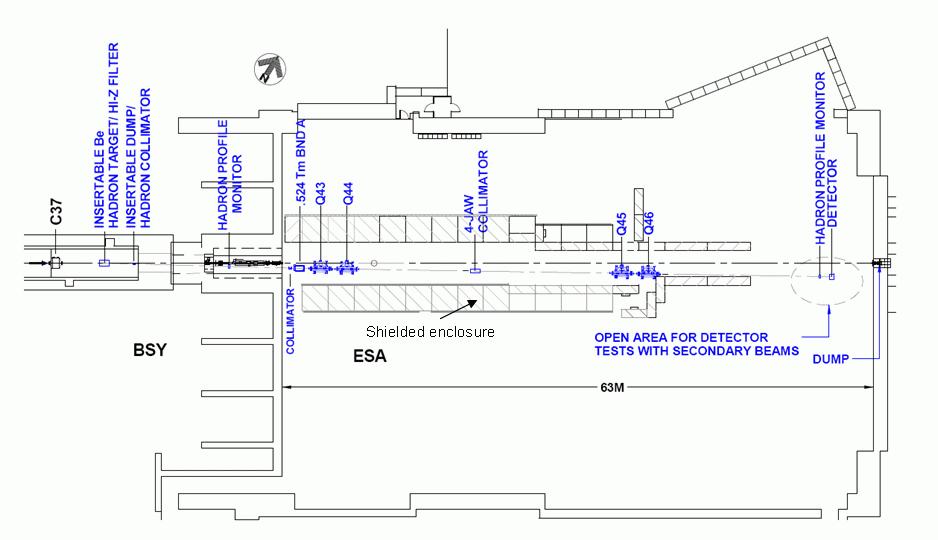}
\caption{\sl End Station A Facility configuration.  Primary beam experiments will be conducted along the primary beam line inside the shielded enclosure.  The primary beam terminates in the beam dump shown in the ESA east wall. Secondary beam tests for detector studies will take place in an open region at the end of ESA.  The proposed hadron beam line components and the new beam dump are shown in blue, overlaid onto the existing ESA setup.}
\label{fig:plan_esa}
\end{center}
\end{figure}

\begin{figure}[htbp]
\begin{center}
\includegraphics[width=0.5\textwidth]{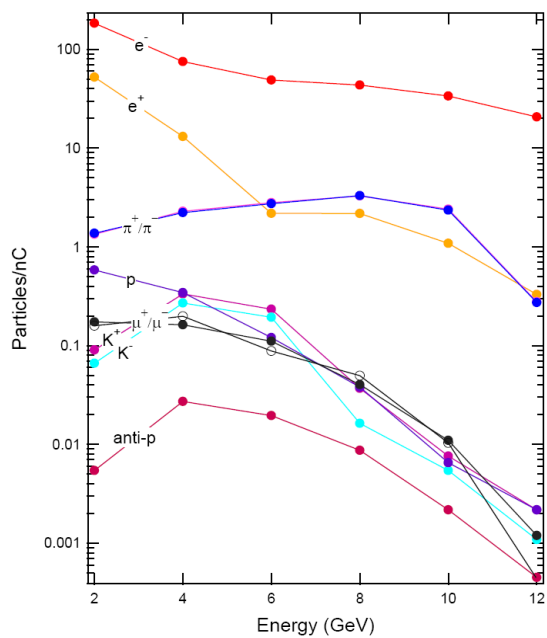}
\caption{\sl Secondary particle yields in ESA per nC of LCLS beam incident on the 0.87 r.l. Be target. The production angle is 1.50 degrees, the acceptance is 5\,${\rm \mu}$sr, and the momentum bite $\Delta p/p = \pm 1\%$.  LCLS beam energy is 13.6\u{GeV}. For expected operating conditions, the yields at the end of ESA are roughly a factor of 4 lower.}
\label{fig:yield_esa}
\end{center}
\end{figure}

End Station Test Beam (ESTB) is a approved and funded SLAC project to use a small fraction of the 13.6\u{GeV} electron beam from 
the Linac Coherent Light Source (LCLS) to restore beam test capabilities in End Station A (ESA), as shown in the schematic  diagram in Figure~\ref{fig:plan_esa}. 
Four new kicker magnets will be installed in the Beam Switch Yard (BSY) to divert 5\u{Hz} of LCLS beam to the 
A-line. This beam can be transported all the way to ESA for beam instrumentation and accelerator physics studies at full electron 
beam intensity. Alternatively, it can be directed against a thin screen in the A-line, to produce secondary electrons or positrons 
with energies up to the incident energy, and a wide range of intensities including single particles per pulse suitable for detector 
studies. The installation of a secondary hadron target and a hadron beam line in ESA is a possible upgrade for 2011. This beam will 
produce pions and kaons over a broad range of momenta, suitable for particle physics and astrophysics detector development or 
calibration in ESA. 
Besides the four new kicker magnets, a new Personnel Protection System (PPS) and a new beam dump in the ESA East wall need to 
be installed. For the hadron target a new beam line with bend and quadrupole magnets and acceptance collimator needs to be 
designed and installed.
The ESTB is a unique resource in all of High Energy Physics for studies requiring high energy, high intensity, low emittance 
electron beams in a large experimental area. These studies include accelerator instrumentation, linear collider accelerator and 
machine-detector interface (MDI) R\&D, development of radiation-hard detectors, material damage studies, and astroparticle detector 
research. As summarised in Table~\ref{tab:slacbeam}, ESTB also provides moderate energy (E=13.6\u{GeV}) secondary beams of electrons and hadrons for 
detector R\&D. Electron beams of exceptional purity, momentum definition, and small size can be delivered. The time structure 
of the test beams is that of the SLAC linac, and is unique in delivering picosecond pulses at known times. This makes triggering 
and data collection very convenient at ESTB. A tagged photon beam could also be provided. At a later stage pions are available 
up to about 12\u{GeV/c} at an intensity of 1 particle/pulse, and kaons at a 1/10 of the pion rate.  
ESTB utilises the existing ESA, a large experimental hall 60 meters in length with 15 and 50-ton overhead cranes and excellent 
availability of utilities, cable plant, and components for mounting experiments. ESA is ideal for detector development and 
testing large scale prototypes or complete systems with high energy particles. Figure~\ref{fig:yield_esa} shows the secondary particle yield per LCLS beam intensity in nC as a function of secondary particle energy.
Funding for the four kicker magnets, new beam dump and a new PPS system is available in early 2010. We have already started 
with designs. The biggest task is the new PPS for ESA, where we expect the completion in early 2011, after which operation 
can commence. Funding for the hadron beam line is expected through 2011.

\begin{table}[htdp]
\begin{center}
\begin{tabular}{@{} |c|c|c| @{}}
 \hline
    Parameters  & BSY & ESA \\
    \hline
    Energy/GeV & 13.6 &  13.6 \\
    \hline
    Repetition rate/Hz& 5 &  5 \\
    \hline
    Charge per pulse/$10^{10}$\,nC& 0.15-0.6 &  0.15-0.6  \\
    \hline
    Energy spread, $\sigma_E/E$ & 0.058\% &  0.058\% \\
    \hline
    Bunch length, rms/$\umu$& 10 &  280  \\
    \hline
    Emittance, rms($\gamma\epsilon_x, \gamma\epsilon_y$)/$10^{-6}$\,mrad& 1.2, 0.7 &  4, 1 \\
    \hline
    Spot size at waist, $\sigma_{x,y}/\umu$ & - &  10\\
    \hline
    Momentum dispersion, $\eta$ and $\eta'$/mm & - & $<10$ \\
    \hline
    Driftspace available & & \\
    for experimental apparatus/m & - & 60\\
    \hline
    Driftspace available & & \\
    for experimental apparatus/m & - & $5\times5$\\
    \hline
\end{tabular}
\end{center}
\caption{\sl ESTB primary electron beam parameters and experimental area at the BSY and in ESA.}
\label{tab:slacbeam}
\end{table}%

\subsection{Asian facilities}
There are several low energy beam test facilities in Asia,
where test of small units can be performed.
\subsubsection{J-PARC}

The 50\u{GeV} proton synchrotron
started its operation at 30\u{GeV} in 2009.
In the hadron physics facility, there are several beam lines.
The K1.1 beam line will be available in 2010,
where hadrons with momentum 0.5$\sim$1.1\u{GeV/c}
and good enough particle yields
are available . This beam line can be used for beam test experiments
until preparation of the main experiment at K1.1 is started.
The K1.8BR beam line is dedicated to the beam test experiments,
and hadrons with momentum 0.5$\sim$1.5\u{GeV/c}
are available.
This beam line also will be ready in 2010.
However, the particle yields are expected to be very low at the beginning
to be used for the experiments.
until the intensity of the proton synchrotron becomes close to the
design value (100\u{MW}).

\subsubsection{KEK}

FTBL (Fuji Test Beam Line)
utilises synchrotron photons radiated from KEKB electron beam
to make electron beams with momentum
0.4$\sim$3.4\u{GeV/c}.
FTBL has been used for many beam test experiments, including ILC activities,
since FTBL started its operation in 2007.
FTBL is not currently available
because of the shutdown (2010$\sim$2012) for the upgrade of KEKB.
ATF (Accelerator Test Facility) for the ILC
can be in principle used for beam test
activities.
The electron beam with momentum 1.4\u{GeV/c} has a bunch structure (2.8\u{ns}).
and the particle yield is $10^{10}$/s.

\subsection{IHEP, Beijing}

BTF (Beijing Testbeam Facility) provides
primary electron beam with momentum 1.1$\sim$1.5\u{GeV/c}
and secondary beams with momentum 0.4$\sim$1.2\u{GeV/c}.
BTF is now under a long shut down (2008-2010) for its upgrade.

\subsection{Tohoku University}

The Research center for electron photon science at Tohoku University in Japan
has a beam test facility providing electrons with momentum 300\u{MeV/c}
and 1.2\u{GeV/c}.
The availability of the facility is very high.

\section{Semi-Permanent beam lines and combined beam tests}
\label{sec:infra}

The establishment of beam lines mainly dedicated to linear collider detector R\&D, called semi-permanent beam lines hereafter, has been an important topic at the workshop. In general it is felt that the establishment of those beam lines would lead to important synergies. This leads from practical issues like ''knowing where the trigger counters are'' to the possibility to install infrastructural components like communication services at the beam test sites. The main advantages of permanent beam lines are listed in the following:
\begin{itemize}
\item The use of a semi-permanent beam line would allow the sharing of experience with the usage of a beam line. 
Hence, the data taking can be much more efficient as the sometimes tedious period of getting up and running can be much shorter;
\item The existence of a semi-permanent beam line would foster the development of common DAQ interfaces which after all would also facilitate the data taking a lot. This can go as far that manning of shifts can be shared by different detector types, simply because the interfaces to the detectors are familiar. This in turn safes travel money and man power. Clearly, it has to be made sure that in particular young students can still be trained at beam test sites;
\item A semi-permanent beam line would facilitate a situation in which one subsystem is the main user while another one
acts as a secondary user to e.g. take calibration data or for long term studies. A general familiarity with a given
beam line would render such a configuration much easier and allows for flexible switches between detector components if circumstances demand it;
\item A common remote control system may allow for data taking even if no expert of a sub-system is on-site. Clearly, this has to comply with safety aspects at the beam test sites;
\item A semi-permanent beam line would naturally lead to a mutual better understanding of other detector components. The fact that a common DAQ system at an early stage may facilitate the system integration in the real detector is also not to be underestimated.
\end{itemize}

In order to underline the need of semi-permanent beam lines, beam requests could be transmitted 
to sites in a coordinated way by the spokespersons of the detector R\&D
collaborations at given dates in a year. By that, several requests from the community arrive at the same 
time which may naturally lead to an assignment of only a few beam lines to the requests. The placing 
of the requests to the sites will be preceded by a brief meeting of the spokespersons in order to have an idea of schedules which could then also be streamlined. The step to a common request is not that long in that case.  A short meeting on coming beam test activities will become a standing item at each linear collider workshop.

All beam test efforts will be monitored by a light monitoring system. In practise, this will be a simple date base where the groups enter the date and the purpose of the test as well as the beam line they use. This is a simple mean to facilitate communication beyond different detector system. It is very light weight and easy to implement at any computing centre (FNAL, DESY, CERN, CC IN2P3). The data base can be brought in operation during the summer/autumn of 2010.

Another question is whether the community should plan for combined beam tests, i.e. combining different detector technologies. The workshop could not identify a clear project of a major combined beam test for the period 2010-2013. There are, however,  occasions at which a combination at a smaller level seems to be feasible. Calorimeters for example need very often a good point resolution. This requirement is very much met by the EUDET Telescope. It could however be imagined that such a task can be realised by a silicon tracking device conceived for linear collider detectors


\section{Concluding remarks}
This document witnesses the large amount of challenging activities in the R\&D for linear collider detectors. 
All proposed technologies need considerable beam test resources in the coming 2--3 years. In view of the DBDs to be completed 
at the end of 2012, a high availability of beam test sites in the coming 2 1/2 years is of major importance. 
This is more important than e.g. the establishment of an ILC beam structure which on the other hand would enhance the validity of
beam test results.
Considerations to shutdown the beam test areas at CERN and FNAL during 2012 bear a considerable 
risk for all the projects. However, the sizable number of sites may allow for searching of alternatives 
in case a given site is not available. This document will support the coordination of the beam test activities of the linear collider
detectors.



\section*{Acknowledgements}
The linear collider detector community would like to thank the LAL directorate for giving us
to opportunity to hold this workshop under optimal working conditions. We would like to thank
in particular the workshop secretaries Val\'erie Brouillard and Patricia Ch\'emali as well as G\'erard Dreneau for the technical support. Thanks to Paul Dauncey for clarifying remarks on the 
status of the DECAL.
\section*{Appendix}
Primary contacts for site managers:\\

\noindent
LC testbeam working group:\\
Kiyotomo Kawagoe (Kobe University) kawagoe@kobe-u.ac.jp\\
Jaehoon Yu (UTA) jaehoonyu@uta.edu\\
Vaclav Vrba (FZU Prague) vrba@fzu.cz\\
Felix Sefkow (DESY) felix.sefkow@desy.de\\

\noindent
Chair of detector R\&D panel:\\
Marcel Demarteau (FNAL) demarteau@fnal.gov\\
\noindent
Editor of this document:\\
Roman P\"oschl (LAL Orsay) poeschl@lal.in2p3.fr\\

These persons may serve as a primary contact in case of additional questions on project plans and
will establish the contact to the various groups. 
\begin{enumerate}
  \small
  \itemsep 0cm \parsep 0cm \topsep 0cm
\item[a.]  
  {\em Hadron and Jet Detection with a Dual-Readout Calorimeter}, {\it NIM} {\bf
    A537} (2005) 537-561.
\item[b.]  
  {\em Electron Detection with a Dual-Readout Calorimeter}, {\it NIM} {\bf A536}
  (2005) 29-51.
\item[c.]  
  {\em Muon Detection with a Dual-Readout Calorimeter}, {\it NIM} {\bf A533}
  (2004) 305-321.
\item[d.]  
  {\em Comparison of High-Energy Electromagnetic Shower Profiles Measured with
  Scintillation and Cerenkov Light}, {\it NIM} {\bf A548} (2005) 336-354.
\item[e.]  
  {\em Separation of Scintillation and Cerenkov Light in an Optical Calorimeter},
  {\it NIM} {\bf A550} (2005) 185-200.
\item[f.]  
  {\em Measurement of the Contribution of Neutrons to Hadron Calorimeter
  Signals}, {\it NIM} {\bf A581} (2007) 643.
\item[g.]  
  {\em Contributions of \v{C}erenkov Light to the Signals from Lead Tungstate
  Crystals}, {\it NIM} {\bf A582} (2007) 474.
\item[h.]  
  {\em Comparison of High-Energy Hadronic Shower Profiles Measured with
  Scintillation and Cerenkov Light}, {\it NIM} {\bf A584} (2008) 273.
\item[i.]  
  {\em Dual-Readout Calorimetry with Lead Tungstate Crystals}, {\it NIM} {\bf
    A584} (2008) 304.
\item[j.]  
  {\em Effects of the Temperature Dependence of the Signals from Lead Tungstate
  Crystals}, {\it NIM} {\bf A593} (2008) 530.
\item[k.]  
  {\em Separation of Crystal Signals into Scintillation and \v{C}erenkov
  Components}, {\it NIM} {\bf A595} (2008) 359.
\item[l.]  
  {\em Neutron Signals for Dual-Readout Calorimetry}, {\it NIM} {\bf A598}(2008)
  422.
\item[m.]  
  {\em Dual-Readout Calorimetry with Crystal Calorimeters}, {\it NIM} {\bf
    A598}(2008) 710.
\item[n.]  
  {\em New crystals for dual-readout calorimetry}, {\it NIM} {\bf A604} (2009)
  512.
\item[o.] 
  {\em Dual-readout calorimetry with a full-size electromagnetic section}, {\it
    NIM} {\bf A610} (2009) 488.
\item[p.]  
  {\em Optimization of crystals for applications in dual-readout calorimetry},
  {\it NIM} {\bf A621} (2010) 212.
\end{enumerate}

\bibitem{FCAL}
  The FCAL Collaboration \\
  \url{http://www-zeuthen.desy.de/ILC/fcal/}

\bibitem{tpc-res}
ILC Global Design Effort and World Wide Study,\\ 
{\em International Linear Collider Reference Design Report},\\ 
ILC-Report-2007-001,\\ 
\url{http://www.linearcollider.org/cms/?pid=1000437}.

\bibitem{mpgd-1}
Y. Giomataris {\it et al.},\\ 
Nucl. Instr. and Meth. A {\bf 376} (1996) 29.

\bibitem{mpgd-2}
F. Sauli,\\ 
Nucl. Instr. and Meth. A {\bf 386} (1997) 531.

\bibitem{npro-1}
LC-TPC collaboration,\\ 
{\em TPC R\&D for an ILC detector},\\ 
LC-DET 2007-005 DESY, Hamburg, Germany. 

\bibitem{npro-2}
LC-TPC collaboration,\\ 
{\em TPC R\&D for a Linear Collider Detector},\\ 
Status report to the DESY PRC65 meeting, April 2008. 

\bibitem{klaus}
K.\,Dehmelt for the LC-TPC Collaboration,\\ 
{\em A Large Prototype of a Time Projection Chamber for a Linear Collider Detector},\\ 
Talk given at TIPP09, Tsukuba, Japan, March 2008 (Submitted to TIPP09 Proceedings in NIMA), 

\bibitem{delent}
G.~W.~P.~De Lentdecker for the LC-TPC Collaboration,\\
{\em A Large TPC Prototype for an ILC Detector},\\ 
Talk given at IEEE Nuclear Science Symposium and Medical Imaging Conference, Orlando, Florida, USA, October 2009.

\bibitem{saltro}
L.~Musa,\\
{\em Prototype compact readout system}.\\
EUDET-Memo-2009-31,\\
\url{http://www.eudet.org/e26/e28/e42441/e79100/EUDET-Memo-2009-31.doc}.


\bibitem{co2cool}
B.~Verlaat and A.~Pieter Colijn,\\ 
{\em CO2 Cooling Developments for HEP Detectors},\\ 
Talk given at VERTEX 2009, Putten, The Netherlands, 13-18 September, 2009.

\bibitem{Roloff:2009zza}
  P.~Roloff,\\
  {\em The EUDET high resolution pixel telescope},\\
  Nucl.\ Instrum.\ Meth.\  A {\bf 604} (2009) 265.

\bibitem{Gregor:2009yy}
  I.~M.~Gregor,\\
  {\em Summary of One Year Operation of the EUDET CMOS Pixel Telescope},\\
  arXiv:0901.0616 [physics.ins-det].

\bibitem{CAPTAN} 
R.~Rivera,\\ 
{\em CAPTAN : Compact and programmable data acquisition node},\\ 
presented at LCTW09, November 2009, Orsay, France.\\
\url{http://ilcagenda.linearcollider.org/conferenceDisplay.py?confId=3735}

\bibitem{Turqueti:2008zz}
  M.~Turqueti, R.~A.~Rivera, A.~Prosser, J.~Andresen and J.~Chramowicz,\\
  {\em Captan: A Hardware Architecture For Integrated Data Acquisition, Control, And Analysis For Detector Development},\\
  FERMILAB-PUB-08-527-CD

\bibitem{Rivera:2009zz}
  R.~A.~Rivera, M.~Turqueti and L.~Uplegger  [USCMS Collaboration],\\
  {\em A Telescope Using CMS PSI46 Pixels and the CAPTAN for Acquisition and Control over Gigabit Ethernet},\\
  FERMILAB-CONF-09-575-CD

\bibitem{alibava}
  R.~Marco-Hernandez and ALIBAVA Collaboration\\
  {\em A portable readout system for microstrip silicon sensors (ALIBAVA)},\\ 
  IEEE Transactions on Nuclear Science, Vol. 56, No. 3, June 2009.





\bibitem{wing-memo}
  M.~Wing,\\ 
  {\em Calorimeter DAQ status},\\ 
  EUDET-Memo-2009-27.\\
  \url{http://www.eudet.org/e26/e28/e42441/e70539/EUDET-MEMO-2009-27.pdf}

\bibitem{decotigny-lctw09}
  D.~Decotigny,\\ 
  {\em CALICE DAQ \& testbeams : status and perspectives},\\ 
  presented at LCTW09, November 2009, Orsay, France.\\
  \url{http://ilcagenda.linearcollider.org/conferenceDisplay.py?confId=3735}

\bibitem{doocs}
  \url{doocs.desy.de}

\bibitem{xdaq}
  \url{https://svnweb.cern.ch/trac/cmsos}

\bibitem{haas-report}
  D.~Haas,\\ 
  {\em The EUDET Pixel Telescope Data Acquisition System},\\ 
  EUDET-Report-2009-03.\\
  \url{http://www.eudet.org/e26/e26/e27/e50991/eudet-report-09-03.pdf}

\bibitem{haas-talk}
  D.~Haas,\\ 
  {\em The EUDAQ Data Acquisition for the JRA1 Pixel Telescope},\\
  presented at LCTW09, November 2009, Orsay, France.\\
  \url{http://ilcagenda.linearcollider.org/conferenceDisplay.py?confId=3735}

\bibitem{killenberg-talk}
  M.~Killenberg,\\ 
  {\em DAQ for LC-TPC},\\ 
  presented at LCTW09, 
  November 2009, Orsay, France.\\
  \url{http://ilcagenda.linearcollider.org/conferenceDisplay.py?confId=3735}

\bibitem{tango}
  \url{http://www.tango-controls.org/}

\bibitem{marlin_et_al}
  F.Gaede, J.Engels,\\ 
  {\em Marlin et al - A Software Framework for ILC detector R\&D},\\
  EUDET-Report-2007-11 \\
  \url{http://www.eudet.org/e26/e27/e584/eudet-report-2007-11.pdf} 

\bibitem{lcio}
  F.~Gaede, T.~Behnke, N.~Graf and T.~Johnson,\\ 
  {\em LCIO: A persistency framework for linear collider simulation studies},\\
  {\it Proceedings of CHEP 03, La Jolla, California, 24-28 Mar 2003, pp TUKT001}

\bibitem{mokka}
  P.~Mora de Freitas and H.~Videau,\\
  {\em Detector simulation with MOKKA GEANT4: Present and future},\\
  {\it Prepared for LCWS 2002, Jeju Island, Korea, 26-30 Aug 2002}

\bibitem{calice_control}
  S.~Karstensen, \\
  {\em Communication Tools},\\ 
  presented at LCTW09, November 2009, Orsay, France.\\
  \url{http://ilcagenda.linearcollider.org/conferenceDisplay.py?confId=3735}

\bibitem{calicesw}
  R.~P\"oschl, N.~Meyer,\\ 
  {\em Calice Software},\\ 
  presented at LCTW09, November 2009, Orsay, France.\\
  \url{http://ilcagenda.linearcollider.org/conferenceDisplay.py?confId=3735}
\bibitem{marlintpc}

  M.~Killenberg, {\em TPC Testbeam Software},\\ 
  presented at LCTW09, November 2009, Orsay, France.\\
  \url{http://ilcagenda.linearcollider.org/conferenceDisplay.py?confId=3735}


\section*{List of participants}
\begin{table}[!h]
\begin{center}
\begin{tabular}{ll}
Name & Institute \\
\hline
Marc Anduze & LLR - \'Ecole Polytechnique/CNRS/IN2P3\\ 
Vincent Boudry & LLR - \'Ecole Polytechnique/CNRS/IN2P3\\  
Jean-Claude Brient & LLR - \'Ecole Polytechnique/CNRS/IN2P3\\  
Alexandre Charpy & LPNHE - Universit\'e Pierre et Marie Curie/CNRS/IN2P3\\ 
Maximilien Chefdeville & LAPP - Universit\'e de Savoie/CNRS/IN2P3\\  
Christophe de La Taille & LAL/OMEGA - Universit\'e Paris XI/CNRS/IN2P3\\  
David Decontigny  & LLR - \'Ecole Polytechnique/CNRS/IN2P3\\
Klaus Dehmelt & DESY/Hamburg\\
Philippe Doublet & LAL - Universit\'e Paris XI/CNRS/IN2P3\\
Gene Fisk & Fermilab\\
Frank Gaede & DESY Hamburg\\
Daniel Haas & Universit\'e de Gen\`eve\\ 
Carsten Hast & SLAC\\
Daniel Jeans & LLR - \'Ecole Polytechnique/CNRS/IN2P3\\ 
Yannis Kariotakis & LAPP - Universit\'e de Savoie/CNRS/IN2P3\\  
Sven Karstensen & DESY Hamburg\\ 
Martin Killenberg & Universit\"at  Bonn\\
Szymon Kulis & AGH-UST \\
Lucie Linssen & CERN\\
Pierre Matricon & LAL - Universit\'e Paris XI/CNRS/IN2P3\\
Takeshi Matsuda & KEK and DESY Hamburg\\
Norbert Meyners & DESY Hamburg\\
Aurore Navoy-Savarro & LPNHE - Universit\'e Pierre et Marie Curie/CNRS/IN2P3\\ 
Giovanni Pauletta & Sezione Di Trieste, Presso l'area Di\\
Roman P\"oschl & LAL - Universit\'e Paris XI/CNRS/IN2P3\\
Martin Pohl & Universit\'e de Gen\`eve\\ 
Erik Ramberg & Fermilab\\
Jos\'e Repond & Argonne National Laboratory\\
Fran{\c c}ois Richard & LAL - Universit\'e Paris XI/CNRS/IN2P3\\
Felix Sefkow & DESY Hamburg\\
Ron Settles & Max Planck Institut f\"ur Physik M\"unchen\\
Petr Sicho & Institute of Physics Prague\\
Tohru Takeshita & Shinshu University\\
Jaap Velthuis & University of Bristol \\
Henri Videau & LLR - \'Ecole Polytechnique/CNRS/IN2P3\\
Marcel Vos & IFIC Valencia\\
Sebastian Weber & Universit\"at Wuppertal\\ 
Vaclav Vrba & Institute of Physics Prague\\ 
Matthew Wing & UCL London\\
Jaehoon Yu & University of Texas at Arlington\\
\end{tabular}
%
\end{center}
\end{table}

%
\end{document}